\documentclass{aa}

\usepackage[utf8]{inputenc}
\usepackage{bm}
\usepackage{algpseudocode}
\usepackage{algorithm}
\usepackage{amsmath}
\usepackage{xcolor}
\usepackage{subfig}
\usepackage{comment}
\usepackage{textgreek}
\DeclareMathOperator*{\argmax}{arg\,max}

\newcommand{\action}{\boldsymbol{a}_t}

\newcommand{\st}{\boldsymbol{s}_t}
\newcommand{\stp}{\boldsymbol{s}_{t+1}}

\usepackage{graphicx}
\usepackage{txfonts}

\begin{document} 

   \title{Focal plane wavefront control with model-based reinforcement learning}

   \subtitle{I. Proof of concept on simulated static and dynamic non-common path aberrations}

 \author{J. Nousiainen
          \inst{1}
          \and I. Taskin \inst{2,1}\fnmsep\thanks{F.R.S.-FNRS FRIA grantee}
          \and M. Kasper\inst{1}
          \and G. Orban de Xivry\inst{2}
          \and O. Absil\inst{2}\fnmsep\thanks{F.R.S.-FNRS Research Director}
          }

   \institute{European Southern Observatory (ESO), Karl-Schwarzschild-Str.\ 2, 85748 Garching, Germany  \\\email{jalo.nousiainen@eso.org} \and 
   STAR Institute, Université de Liège, Allée du Six Août 19C, 4000 Liège, Belgium
             }

   \date{Received 8 December, 2025; accepted 31 March, 2026}

  \abstract
   {The direct imaging of potentially habitable exoplanets is one prime science case for high-contrast imaging (HCI) instruments on ground-based, extremely large telescopes. Most such exoplanets orbit close to their host stars, where their observation is limited by fast-moving atmospheric speckles and quasi-static non-common path aberrations (NCPA).}
   {Conventional NCPA correction methods often use mechanical mirror probes, which compromise performance during operation. This work presents machine-learning-based NCPA control methods that automatically detect and correct both dynamic and static NCPA errors by leveraging past telemetry data and sequential phase diversity.}
   {We extend previous work in reinforcement learning (RL) for Adaptive optics (AO) to focal plane wavefront control. A new model‑based RL algorithm, Policy Optimization for Non‑common Path Aberrations (PO4NCPA), interprets the focal‑plane image as input data and, through sequential phase diversity, determines phase corrections that optimize both non‑coronagraphic and post‑coronagraphic point‑spread functions (PSFs) without prior system knowledge. Furthermore, we demonstrate the effectiveness of this approach by numerically simulating static NCPA errors on a ground-based telescope and an infrared imager affected by water-vapor-induced seeing (dynamic NCPAs).
   }
   {Simulations show that PO4NCPA robustly compensates static and dynamic NCPAs. In static cases, it achieves near-optimal focal-plane light suppression with a coronagraph and near-optimal Strehl without one. With dynamics NCPA, it matches the performance of the modal least-squares reconstruction combined with a 1-step delay integrator in these metrics, though with a higher wavefront RMS, especially for high-order modes. The method remains effective for the ELT pupil, vector vortex coronagraph, and under photon and background noise.  
   }
   {PO4NCPA is model-free and can be directly applied to standard imaging as well as to any type of coronagraphy. Its sub-millisecond inference times and performance also make it suitable for real-time low-order correction of atmospheric turbulence beyond HCI requirements.}
   
   \keywords{Instrumentation: adaptive optics, high angular resolution  -- Methods: data analysis, numerical -- Techniques: high angular resolution} 
   \maketitle
%

\section{Introduction}
Studying extrasolar planets (exoplanets) and their systems is one of the fastest-growing areas in modern astrophysics. To date, over 6,000 confirmed exoplanets have been discovered, primarily using indirect methods such as radial velocity and photometric transit observations\footnote{Exoplanet Orbit Database: http://exoplanets.org/}. High-contrast imaging (HCI) aims to separate exoplanet light from stellar light optically, thereby allowing direct characterization of the exoplanet light. However, HCI observations have only managed to detect a few tens of young, luminous giant exoplanets (e.g., \citealp{2023ASPC..534..799C}). This is due to the extreme contrast required to observe exoplanets located a fraction of an arcsecond from their host stars, which can be up to a billion times brighter than the planets themselves.

For ground-based observations, HCI combines eXtreme Adaptive Optics (XAO, e.g., \citealp{guyon2005limits, guyon2018extreme}) and coronagraphy \citep{guyon2018extreme} with a way to distinguish stellar speckles produced by imperfect instrument optics and atmospheric residual from the exoplanet, such as spectral and angular differential imaging (SDI, ADI;  \citealp{2004ApJ...615L..61M, 2006ApJ...641..556M}) or high-dispersion spectroscopy \citep{2015A&A...576A..59S}. 
The dominant noise sources for these approaches are the XAO residual halo \citep{guyon2018extreme,2021A&A...646A.150O} caused by an imperfect control of the adaptive optics system, and the non-common path (NCPA) errors (not seen by the AO system) caused by imperfect optics, slowly evolving temperature changes, wavefront discontinuities, changes in the gravity vector and/or a chromaticity mismatch between the wavefront-sensing wavelength of the XAO system and the science camera wavelength. 

HCI performance can suffer from three types of NCPA errors: static, quasi-static, and dynamic. Static NCPA errors create static speckles on the focal plane, which can usually be calibrated with the instrument's internal light source and are generally easier to post-process. Quasi-static speckles evolve slowly (compared to AO corrected atmospheric speckles) over time; they do not average out during typical observation periods. ADI and SDI struggle to effectively remove quasi-static speckles at small angular separations because such speckles move slowly with field rotation or wavelength. One important class of quasi-static wavefront aberrations that produce slowly varying speckles and severely limit the contrast performance for HCI instruments is the so-called low wind effect (LWE). The LWE effect is due to the radiative cooling of the telescope spiders (support structure of the secondary mirror), which creates air temperature inhomogeneities that appear as phase discontinuities. These discontinuities are not properly detected by some WFSs (e.g., modulated pyramid and Shack-Hartmann sensor) and are therefore poorly corrected by the AO system. LWE has been reported as the limiting error term under low wind-speed conditions with several HCI instruments (e.g., SPHERE, \citealp{milli2018low}; SCExAO \citealp{jovanovic2015subaru} and GPI, \citealp{macintosh2008gemini}).
Dynamic NCPAs, which tend to average out, create additional photon noise and boost speckle noise through speckle coupling. Compensating for dynamic speckles is particularly important when their contribution is significant compared to XAO residuals, as is the case, for example, with the water vapor (WV) seeing \citep{absil2022impact} that plagues N-band observations of the Mid-infrared ELT Imager and Spectrograph (METIS, \citealp{brandl2024final}). While the refractive index of dry air is nearly achromatic across the visible and infrared ranges, the chromaticity of the water vapor refractive index introduces a seeing component in N-band that is not seen by the AO WFS operating at optical or near-infrared wavelengths.

One way to address the quasi-static and dynamic NCPA problem is to apply active correction at the hardware level, for example, by modifying the AO system's flat reference or by controlling a separate NCPA deformable mirror (DM) in the science camera path. As the AO WFS does not measure them, NCPA errors must be observed either from the focal plane with a focal-plane wavefront sensor (FPWFS) or with a WFS placed in the science camera's optical path, for example, using a reflective coronagraph mask \citep{guyon2009coronagraphic, singh2014lyot}. This paper focuses on the FPWFS case and assumes a separate NCPA DM to be available in the science camera.

One difficulty with FPWFS lies in the observation model: the focal-plane pupil-plane relationship is nonlinear and degenerate \citep{guyon2018extreme}, and this is even more pronounced when combined with different coronagraph designs. Most importantly, recovering the NCPA errors from a single focal-plane image is an ill-posed problem, because two different phase patterns in the pupil plane can produce the same focal-plane image. This property, often referred to as phase ambiguity in the literature, can be overcome by including an additional focal-plane image with a known phase offset, such as defocus \citep{1982OptEn..21..829G}. However, this reduces the observing time by reserving time or a portion of the beam solely for wavefront measurements. There are multiple other ways to lift the phase ambiguity, for example, introducing asymmetries in the pupil \citep{martinache2013asymmetric} or Lyot \citep{de2024alf} planes, splitting polarization in a vector vortex coronagraph \citep[VVC,][]{riaud2012instantaneous}, or using scalar vortex coronagraphs \citep{2022Quesnel, de2024vortex}. The phase ambiguity can also be resolved from past data, namely the past NCPA corrections and the corresponding focal plane images, provided the past corrections are non-zero. This approach for lifting phase ambiguity is often referred to as sequential phase diversity \citep{gonsalves2010sequential, keller2012extremely}.

On the other hand, the application of machine learning techniques for wavefront sensing and control in HCI has been an active area of research in recent years. A wide range of methods have been tested in numerical simulations, test-bench setups, and also on-sky (e.g., for XAO by \citealp{van2022predictive} and \citealp{landman2025making}). These methods include supervised learning-based solutions using neural networks (NN), linear models, frequency-based models, as well as deep reinforcement learning (RL) methods. These machine learning-based methods have been shown to improve HCI instruments in many crucial aspects, such as predictive control (e.g., \citealp{nousiainen2021adaptive, pou2024integrating, guyon2017adaptive, dinis2024upgrading}), mitigating WFS non-linearities (e.g., \citealp{landman2024making, 2023Wong, nousiainen2022toward}), FPWFS (e.g., \citealp{2021Orban, 2022Quesnel,terreri2022neural}), and PSF reconstruction (e.g., \citealp{kuznetsov2023prediction}). This paper aims to adapt the policy-optimization RL method introduced for AO control \citep{nousiainen2022toward, nousiainen2024laboratory} to focal-plane wavefront control using sequential phase diversity.  

The paper’s structure is as follows: In Section 2, we introduce the existing literature and position our work in relation to these. Section 3 discusses the prerequisites, providing a short introduction to RL for readers from an AO background, as well as a brief description of focal-plane data for readers with a background in ML. In Section 4, we introduce the control algorithm, the Policy Optimization for NCPA (PO4NCPA). It contains definitions of the state space and control actions/signals, as well as a description of the entire algorithm. Section 5 demonstrates the performance and analyzes the behaviour of PO4NCPA through numerical simulations. Here, we describe the numerical simulations and experiments and visualize the results. The last section (Sec. \ref{sec:discussion}) discusses the findings, future work, and perspectives. 

\section{Related work}
Model-based RL (e.g., \citealp{deisenroth2011pilco, chua2018deep}) has not previously been combined with sequential phase diversity for FPWFS. The present work addresses that gap by adapting the model-based RL framework PO4AO \citep{nousiainen2022toward, nousiainen2024laboratory}, initially developed for adaptive optics (AO) control, to FPWFS with sequential phase diversity. The proposed method, PO4NCPA, follows the same principle of direct policy (neural-network controller) optimization via learned system dynamics (a neural-network model of the optical path), while redesigning the network structure, hyperparameters, and training settings to meet FPWFS requirements.

This study builds on sequential phase-diversity techniques \citep{1982OptEn..21..829G}, which resolve phase-sign ambiguity by exploiting temporal information in focal-plane images. Unlike classical phase diversity, no predefined (e.g., defocused images) are required, eliminating associated observing overheads. The Fast and Furious algorithm \citep{Korkiakoski:14}, relying on this concept, applies linear approximations to iteratively minimize focal-plane aberrations and has been validated on-sky \citep{2020Bos}, though it is incompatible with coronagraphic imaging. The ``2 Fast 2 Furious'' extension \citep{2023Bottom} partially addresses this but is limited to symmetric coronagraphs and excludes complex designs such as vortex coronagraphs. To generalize further, \cite{2023Bottom} proposed ``Tokyo Drift'', a data-driven approach trained with supervised learning on large simulated datasets to handle diverse coronagraph types.

Electric field conjugation (EFC, \citealp{give2007broadband}) is a subclass of FPWFS (with a coronagraph), in which the controller aims not only to minimize phase errors but to minimize the focal-plane electric field intensity in a region of interest (dark hole). EFC typically employs a linear electric-field reconstruction using predefined DM probes (pair-wise probing; \citealp{give2011pair}), which introduces model dependence. Data-driven variants \citep{ruffio2022non, haffert2023implicit} mitigate this by operating directly on focal-plane intensity responses without explicit electric field reconstruction, though they still rely on DM probing. Similar to these methods, PO4NCPA operates directly on DM commands and the corresponding focal-plane image and can enhance post-coronagraphic contrast beyond phase-error correction (with a suitable reward function), but without predefined probes.

Further, both supervised ML and model-free RL have been studied for FPWFS (i.e., wavefront-sensorless AO) beyond sequential phase diversity for recovering phase errors. One of the first uses of NNs in FPWFS used a pair of out-of-focus and in-focus PSFs to predict phase errors and recover a nearly diffraction-limited image \citep{1990Angel}, leveraging supervised learning. Since then, NN-based supervised learning methods have evolved in many ways. For example, more recent papers use more complicated NN, such as U-Nets and ResNets  \citep{2021Orban}, different types of phase diversity, such as phase provided by a vortex coronagraph  \citep{2022Quesnel}, and combining NN with different data preprocessing steps, such as principal component analysis \citep{terreri2022neural}. Moreover,  a pair of out-of-focus and in-focus PSFs has been used in an unsupervised learning scheme with autoencoder NNs \citep{quesnel2022simulator}.  Model-free RL for FPWFS has been demonstrated using in/out-of-focus PSFs as phase diversity \citep{2024Guiterrez}, and explored in broader wavefront-sensorless AO contexts. Examples include deep RL for aberration correction framed as a Markov decision process \citep{ke2019self}, adaptive microscopy using deep RL with deformable mirrors \citep{durech2021wavefront}, and fiber-coupled optical communications using RL-based AO \citep{Parvizi2023}. Unlike these model-free RL approaches, the present work employs model-based RL tightly integrated with sequential phase diversity, focusing on metrics relevant to astronomical high-contrast imaging.

Finally, an alternative direction synchronizes FPWFS with AO wavefront sensors. The DrWHO algorithm \citep{2022Skaf} exemplifies a fully model-free FPWFS approach that combines PyWFS and focal-plane data to correct for slow and static aberrations. Although effective, further improvements on the algorithm performance are needed to be compatible with HCI.

\section{Preliminaries}
\subsection{Reinforcement learning applied to focal plane wavefront control}
\label{sec:prelim}

We begin by introducing some standard notation and terminology commonly used in Reinforcement Learning (RL). A typical framework for formalizing RL problems is the Markov Decision Process (MDP). An MDP is a discrete-time stochastic process in which, at each time step $t$, the system occupies a state $\st \in \mathcal{S}$, where $\mathcal{S}$ denotes the set of all possible states. A decision-maker (or agent) selects an action $\action \in \mathcal{A}$, with $\mathcal{A}$ being the action space, based on the current state. In response, the environment transitions to a new state $\stp$. Since the transition dynamics are stochastic—due, for instance, to evolving stochastic turbulence—they are defined by a conditional probability distribution $p(\stp \mid \st, \action)$.
\footnote{The initial state $\boldsymbol{s}_0$ is sampled from an initial state distribution $p_0(\boldsymbol{s}_0)$.}

At each time step, a reward $R_t = r(\st, \action)$ (a function of the state and action) is observed. The user typically crafts the reward function to encourage the desired behavior of the agent—for example, sharpening the focal-plane PSF or enhancing post-coronagraphic contrast. The actions our decision-maker takes are determined by a ``policy'' $\pi_\xi : \st \mapsto \action$, which is a function that maps states into actions. The objective of reinforcement learning is to find a policy that maximizes the cumulative reward in the given environment governed by the transition dynamics, that is,  
\begin{align}
    \label{eq:obj_of_rl}
    \argmax_\xi \mathbb{E}_{p_\xi({\bf s}_0,..., {\bf s}_T)} \left[ \ \sum_{t=0}^T r(\bm s_t,\pi_\xi(\bm s_{t})) \ \right],
\end{align}
where $\mathbb E$ is the expected value and
\begin{equation*}
p_\xi({\bf s}_0,..., {\bf s}_T) = p_0(\bm s_0) \prod_{t=1}^T p(\st | {\bf s}_{t-1}, \pi_\xi({\bf s}_{t-1})),
\end{equation*}
with the initial distribution $\bm s_0 \sim p_0$ and convention $\pi_\xi({\bf s}_{-1}) = \bm a_0$ for a fixed initial DM commands $ \bm a_0$. In particular, we focus here on parametric models of $\pi_\xi$, where $\xi$ denotes the policy's parameter set, namely the weights and biases of a NN. That is, given that the actions are given by $\pi_\xi$, we wish to find the parameters $\xi$ that maximize the expected cumulative reward the decision-maker receives. Here $T$ is the maximum length of an episode or a single run of the algorithm in the environment.

The assumption of RL is that the transition dynamics is not known: it includes a multitude of unknowns, including the stochastically moving NCPAs, LWE, imperfections in the coronagraph, and other optical elements. In order to solve Eq. \eqref{eq:obj_of_rl} efficiently, model-based RL algorithms estimate the true dynamics model $p(s_{t+1}|s_t, a_t)$ in \eqref{eq:obj_of_rl} by an approximate model $\hat{p}(s_{t+1}|s_t, a_t)$ (learned from the data). Model-free methods, in turn, only learn a policy -- they do not attempt to model the environment. 

Finally, it is common in RL to use reward functions that are not differentiable (e.g., 1 for winning a game, 0 otherwise) or that do not depend directly on the state. However, the choice of algorithm restricts us to using differentiable reward functions that are directly observed from the state model. An example of such a reward function is an Euclidean distance from an ideal (non-distorted) PSF or the amount of light in the post-coronagraphic PSF.

\subsection{Focal plane wavefront sensing}
This section describes the optical focal-plane models discussed in this paper. We consider three types of focal-plane images (serving as the input to the RL decision-maker): non-coronagraphic (i.e., standard imaging, SI), perfect-coronagraphic (PC), and vector vortex coronagraph (VVC). 

Let us denote the incoming electromagnetic field (the wavefront) on the pupil plane by $\psi:\mathbb{R}^2 \to \mathbb{C}$, defined by:
\begin{equation}\label{eq:frau}
\psi(x,y)  =   A_{\Omega}(x,y)e^{i\phi(x,y)},
\end{equation}
where $A_{\Omega}(x,y)$ is the amplitude over the pupil aperture and $\phi(x,y)$ the phase aberrations. The Fraunhofer approximation of a monochromatic point spread function (PSF) is given by, 
\begin{equation} \label{eq:frau_phi}
s_\phi(u,v) = \vert \mathcal F \{A_{\Omega}e^{i\phi}\}  (u,v) \vert ^2,
\end{equation}
where $\mathcal F\{ .\}$ is a fourier operator. The Eq. \eqref{eq:frau_phi} above describes the focal-plane image we use in our simulations for the non-coronagraphic case (i.e., standard imaging).

Further, we consider two types of coronagraphs: A perfect coronagraph and VVC. The fundamental principle of both coronagraphs is the same – the coronagraph suppresses light from an on-axis source while preserving the off-axis companions’ signal (e.g., an exoplanet circling a host star).
 
A theoretical perfect coronagraph model suppresses all light for an on-axis flat wavefront while preserving the off-axis source \citep{cavarroc2006fundamental, por2018high}. The complex wavefront $\psi$ after this ideal coronagraph on the pupil plane is given by
\begin{equation}\label{eq:coro}
\psi_{0}(x,y)  =  A_{\Omega}(x,y)\left(\exp[i\phi(x,y)] -\sqrt{E_c}\right),
\end{equation}
where $E_c = \exp(-\sigma^2_\phi)$ is the instantaneous coherent energy (also referred to as the Strehl ratio under the Maréchal approximation) and $\sigma_\phi^2$ the spatial variance of the wavefront aberrations. The focal-plane image follows from the subsequent Fraunhofer approximation in Eq.\eqref{eq:frau_phi}. In this paper, we use a circular pupil without a central obstruction for both standard imaging and perfect coronagraphic images to minimize the effects of pupil sampling, non-symmetric spiders, and numerical errors on phase ambiguity and to ensure that our RL framework relies solely on sequential phase diversity to lift any ambiguity in the focal plane. 

\begin{figure}
\includegraphics[trim={0cm 0cm 0cm 0cm}, clip, width=0.47\textwidth]{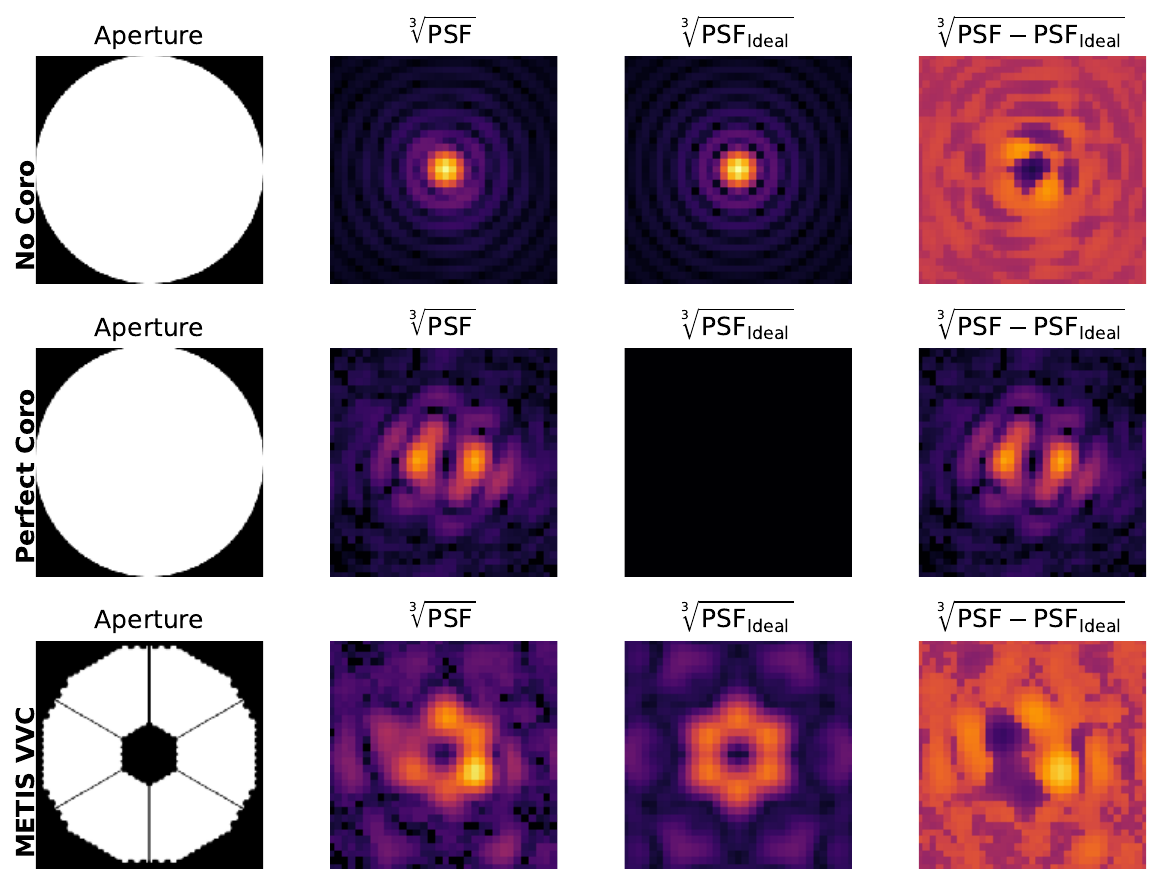}%
\caption{Illustration of the preprocessing step of the focal plane data, i.e, the observation. For a non-coronagraphic system, the Airy pattern (perfect PSF) is subtracted from the PSF, and for the perfect coronagraph, the perfect PSF is simply a dark image (background). The images are then flattened with the cubic root.} 
\label{fig:obs}
\end{figure}

A vector vortex coronagraph combines a vortex phase mask in an intermediate focal plane with a downstream Lyot stop in the pupil plane. The vector vortex phase mask introduces an azimuthal phase ramp on the incoming wavefront, described mathematically as $\exp(\pm i l_p \theta)$, where $\theta$ represents the azimuthal coordinate, $l_p$ is the topological charge of the vortex, and the $\pm$ signs reflect the mask's vectorial nature. This dual-signature arises from the mask's implementation as a half-wave plate with a spatially varying fast axis, creating conjugated phase ramps for the two circular polarization states. The textbook effect of the vortex coronagraph is to move the light of an on-axis source outside the downstream geometric pupil. This diffracted light is then blocked by the Lyot stop $M_{\Omega}$, theoretically achieving perfect starlight cancellation for a circular, unobstructed entrance aperture \cite[e.g.,][]{mawet2005annular}.
Under the Fraunhofer approximation and assuming only phase aberrations in the entrance pupil plane, the monochromatic post-coronagraphic PSF for a vector vortex coronagraph can be expressed as:
\begin{equation}\label{eq:vvc}
s_{\phi}^{\pm}(u,v) = \vert \mathcal F\{M_{\Omega}\mathcal F^{-1} \{e^{\pm i l_p \theta}\mathcal F \{A_{\Omega}e^{i\phi(x,y)}\}\}\} (u,v)\vert ^2
\end{equation}
where the final image for a VVC is obtained by summing the two conjugated PSFs, $s_{\phi}^+$ and $s_{\phi}^-$,  corresponding to the two circular polarization states.
Here, we focus on the most commonly used topological charge, $l_p=2$ \cite[e.g.,][]{Absil+2016}. In numerical experiments, we combine the VVC with the ELT pupil (see Fig.~\ref{fig:obs}). Therefore, the theoretical perfect starlight cancellation is not achieved under a non-aberrated wavefront, but a diffraction pattern is observed.

For all the models discussed above (Eq. \eqref{eq:frau_phi}, \eqref{eq:coro}, and \eqref{eq:vvc}), the Fourier relationship between the focal-plane image and the pupil-plane phase aberration introduces a sign ambiguity for even radial-order Zernike modes \cite[e.g.,][]{martinache2013asymmetric}. That is,
\begin{equation}
    |\mathcal{F}(E_{\mathrm{even}}(x,y))|^{2}
    = 
    |\mathcal{F}(E_{\mathrm{even}}^{*}(-x,-y))|^{2},
    \label{eq:even_ambiguity}
\end{equation}
where $E_{\mathrm{even}}(x,y) = \exp\!\left(-i\phi_{\mathrm{even}}(x,y)\right)$ is the pupil-plane electric field with phase aberrations $\phi_{\mathrm{even}}$ (containing even modes only), and $E_{\mathrm{even}}^{*}$ is its conjugate. In the absence of a diffraction pattern, that is, for the case of a perfect coronagraph, the sign ambiguity extends to both even and odd phase aberrations. More precisely, the perfect coronagraph PSF matches the PSD of the phase assuming small aberration approximation (2nd Taylor Expansion, e.g., \citealp{10.1117/1.JATIS.4.1.019001}), which is ambiguous for sines (even modes) and cosines (odds). A single focal-plane image is not sensitive to the sign of any (even or odd) mode. Moreover, in the case of an ELT pupil (with VVC), one slightly thicker spider\footnote{this slightly thicker spider is due to the shadow of the ELT-M1 crane, located along one of the spiders} in the ELT exit pupil breaks pupil symmetry and reduces the phase ambiguity to a certain extent. Circular pupil (SI and PC) examples to demonstrate the principle of our method, while the ELT-pupil with VVC illustrates a more realistic use case of the algorithm.

\section{Policy optimization for NCPA}
This section describes the control algorithm and the optimization procedures for both the dynamics model \( p_\omega(\mathbf{s}_t, \mathbf{a}_t) \) and the policy \( \pi_\xi(\mathbf{a}_t | \mathbf{s}_t) \). The central concept is to train a dynamics model capable of predicting the next focal-plane image—referred to as the observation—based on prior images (observations) and the applied deformable mirror (DM) commands. This learned model is then used to improve the control policy. The algorithm proceeds iteratively through three steps: 

\begin{enumerate}
    \item Running the policy:  The policy runs NCPA control loop for $T$ timesteps (a single episode).
    \item Improving the dynamics model: the dynamics model is optimized using a supervised learning objective Eq. \eqref{eq:relative_loss}.
    \item Improving the policy: the policy is optimized by using the dynamics model, see Eq.\eqref{eq:optimization_of_hatr}.
\end{enumerate}
During each iteration (steps 1-3), the algorithm gathers one episode of data—consisting of 20 consecutive focal-plane images (time steps) and the corresponding deformable mirror (DM) commands—by executing the policy within the NCPA control loop. The collected observations and actions are stored, and the policy and dynamics model are then updated using gradient-based optimization over all accumulated data. Our experiments (Sec. \ref{sec:results}) show that PO4NCPA converges within around 10-15 time steps for each test case; hence, an episode length of 20 is a good value for learning the full range of needed actions.

The following subsections describe the observation representation, neural network architectures of the dynamics and policy models, and the optimization procedure. 

\subsection{Focal plane wavefront control as a Markov decision process}
\label{sec:state_repre}
The environment in which the RL algorithm operates consists of the electric field of the incoming light, the optics, the potential coronagraph, the phase modification due to NCPA errors, and the deformable mirror. The RL algorithm interacts with the environment by giving residual commands to the DM. Together, these form an MDP, where the next state depends only on the previous state (previous DM position, electric field, and NCPA error) and the DM command applied. However, we do not directly observe the electric field; we measure it with a science camera, with or without a coronagraph, i.e., we measure the intensity of the focal-plane electric field (see Eqs. \ref{eq:frau_phi}, \ref {eq:coro}, and \ref{eq:vvc}). As discussed before, the full electric field cannot be recovered from a single focal-plane image (phase ambiguity). Hence, the focal-plane image can be considered only a partial observation of the underlying MDP. One way to resolve the phase ambiguity in the observation is to include the previous focal-plane image and the given differential DM command, that is, via sequential phase diversity \citep{2002Gonsalves}.

Moreover, we modify the focal plane image to better suit NNs trained with stochastic gradient descent. In the case of non-coronagraphic PSFs and VVC with ELT pupil, a significant amount of light is contributed to the diffraction pattern, noted as $\mathrm{PSF_{ideal}}$ (see the third column in Fig.~\ref{fig:obs}). The focal plane intensity is very unevenly distributed, and the speckles due to optical aberrations closer to the center are up to several hundred times brighter than the speckles further away. Hence, we subtract the diffraction pattern and flatten the resulting image using a cubic root; the final observation is given by (see Fig.~\ref{fig:obs}):
\begin{equation}
\bm o_t = \sqrt[3]{\mathrm{PSF}^t - \mathrm{PSF}_{\rm ideal}^t}.
\end{equation} 
Further discussion on this choice can be found in Sec. \ref{sec:discussion}. The actions are defined as a vector of Zernike mode coefficients applied on top of the last full DM command:  $\bm a_t = (a_t^1, a_t^2, \hdots, a_t^N)$. To prevent the algorithm from learning to steer the light outside the defined focal plane, the deformable mirror is numerically clipped to the maximum expected NCPA error, that is, each mode at three standard deviations of the NCPA spectrum (this parameter is just a crude limit for the clipping, and the algorithm was not sensitive to the exact value). The DM command (action vector) is also normalized with this value. Let $\mathbf{z}_t = (z_{\text{max}}^1,z_{\text{max}}^2, \hdots, z_{\text{max}}^N)$ be the vector of expected max values. The residual command sent to the DM is then given by 

\begin{equation}
\mathbf{z}_t =
\begin{pmatrix}
z_t^1 \\
z_t^2 \\
\vdots \\
z_t^N
\end{pmatrix}
=
\begin{pmatrix}
a_t^1 \, z_{\text{max}}^1 \\
a_t^2 \, z_{\text{max}}^2 \\
\vdots \\
a_t^N \, z_{\text{max}}^N
\end{pmatrix} .
\end{equation}

Now we define the state of the environment (as MDP) by:
\begin{equation}
\bm s_t = (\bm o_{t}, \bm o_{t-1}, \bm a_{t-1}).
\end{equation}
This formulation follows approximately Markovian statistics (previous image and command providing the needed phase diversity), that is, the next state depends (and can be inferred) only on the last state information and the given action $\bm a_{t}$. 

For a state-action pair, the reward was chosen as a negative Euclidean norm between the expected ideal, non-distorted PSF, and observation, that is,
\begin{equation}
r(\bm s_t, \bm a_t) = -{\mathbb E}_{p(\bm s_{t+1} | \bm s_t, \bm a_t)} \| \tilde{\bm o}_{t+1} \|^2,
\end{equation}
where $\tilde{\bm o}_{t+1}$ is a distribution obtained from $\tilde{\bm s}_{t+1} \sim p(\cdot| s_t, a_t)$ (probabilistic transitions dynamics).  The reward here is calculated for the full focal-plane image. It could also be calculated on photometric apertures tailored to the application, for example, a one-sided dark hole. Moreover, because the best reward is achieved when a diffraction-limited PSF is observed, the RL algorithm will never try to ``apodize'' the PSF. In particular, when digging deep, dark holes, the reward (or the ideal PSF subtraction in the observation) must be modified to enable RL to go beyond the perfect PSF via apodization.

\subsection{The dynamics model}
\label{sec:spatial}
\begin{figure}
\includegraphics[clip, width=0.49\textwidth]{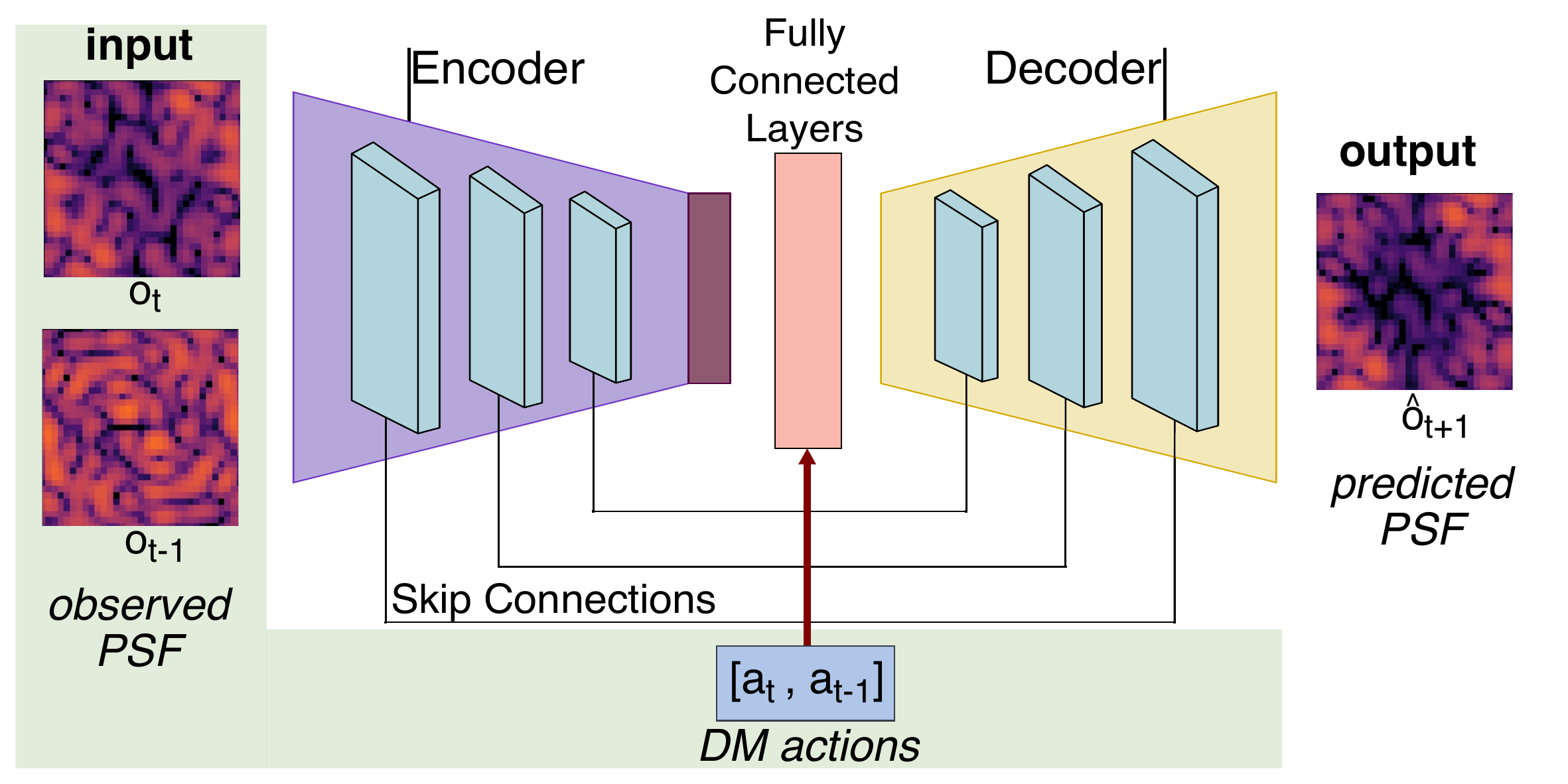}%
\caption{Dynamics model NN design. Trained on closed-loop data (science camera images and residual commands), the dynamics model learns to simulate the optical path of the light. Highlighted in green are the inputs to the NN.} 
\label{fig:dynamics}
\end{figure}

The state information includes two different data tensor shapes. The focal-plane images are 2D, whereas the DM commands are vectors of Zernike modes. We designed the dynamics model to account for the different shapes. The dynamics model has two input channels: one for focal plane images and one for DM commands. The image channel consists of multiple convolutional layers, a flattening operation, and a fully connected layer. The DM command channel concatenates the vector inputs with the output of the images channel vector. The concatenated vector is then propagated through one fully connected layer, which is subsequently reshaped into a 3D tensor. The 3D tensor is then propagated through multiple convolutional layers to form an image representing the next focal plane. The convolutional layers have skip connections to the previous layers (see Fig.~\ref{fig:dynamics}).
This deterministic dynamics model $\hat{p}_{\omega}(\st, \action)$ predicts the next state $\stp$ based on the current state and action. The model parameters $\omega$—representing the neural network’s weights and biases—are learned by executing the policy $\pi$ in the environment, that is, by using the policy to control the AO system, collecting tuples $(\st, \action, \stp)$ into a dataset $\mathcal{D}$, and minimizing the loss function $J$, the root mean square error between the predicted and actual next states (overloading notation for clarity):

\begin{equation}
\label{eq:dynamics_error1}
    J= \sqrt{\sum_{\mathcal{D}} \left\|\stp - \hat{p}_\omega(\st, \action)\right\|^2}
    = \sqrt{\sum_{\mathcal{D}} \left\|\bm o_{t+1} - \hat{\bm o}_{t+1}\right\|^2},
\end{equation}
where ${\bm o}_{t+1}$ is obtained from the state $\stp$ and $\hat{\bm o}_{t+1}$ is the observation predicted by $\hat{p}_\omega(\st, \action)$. 

In focal plane control, the dynamics model must be accurate across a wide range of input strengths. The mean squared error (eq. \ref{eq:dynamics_error1}) takes the square of the absolute error, which effectively causes the optimizer to focus more on wavefronts with relatively larger input values (large wavefront errors, i.e., small Strehl). As a remedy, we follow the approach of \citet{Landman:20} and introduce a relative loss function, $J_{\text{relative}}$ for dynamics optimization, that is, we opted to weigh the mean squared error by the square of the RMS of the true (label) observation:

\begin{equation}
J_{\text{relative}} = 
\left\langle 
\frac{
\sqrt{\sum_{\mathcal{D}} \| \bm{o}_{t+1} - \hat{\bm{o}}_{t+1} \|^2}
}{
\sqrt{\sum_{\mathcal{D}} \| \bm{o}_{t+1} \|^2 + \epsilon}
}
\right\rangle ,
\label{eq:relative_loss}
\end{equation}
where $\langle \cdot \rangle$ denotes the mean over a sampled batch. The $\epsilon$ is introduced to avoid divergence for very small input RMS ($\approx 10^{-7}$). The optimization was done using the Adam algorithm \citep{kingma2014adam}.

It is well established that model-based RL can suffer from poor performance due to overfitting of the dynamics model, which may be overly exploited during control tasks such as planning or policy optimization—particularly in the early stages of training \citep{nagabandi2018neural}. To mitigate this issue, we use an ensemble of dynamics models, where each model is trained on a different bootstrap dataset, that is, a randomly sampled subset of the collected observations. As a result, each model in the ensemble is exposed to a distinct portion of the training data, leading to diverse neural network approximations. During policy training, predictions from all ensemble members are averaged (see lines 13 and 15 of Algorithm~\ref{alg:mbpo}). For further discussion on the use of ensemble models in this context, see \cite{chua2018deep}.

\begin{figure}
\includegraphics[ clip, width=0.49\textwidth]{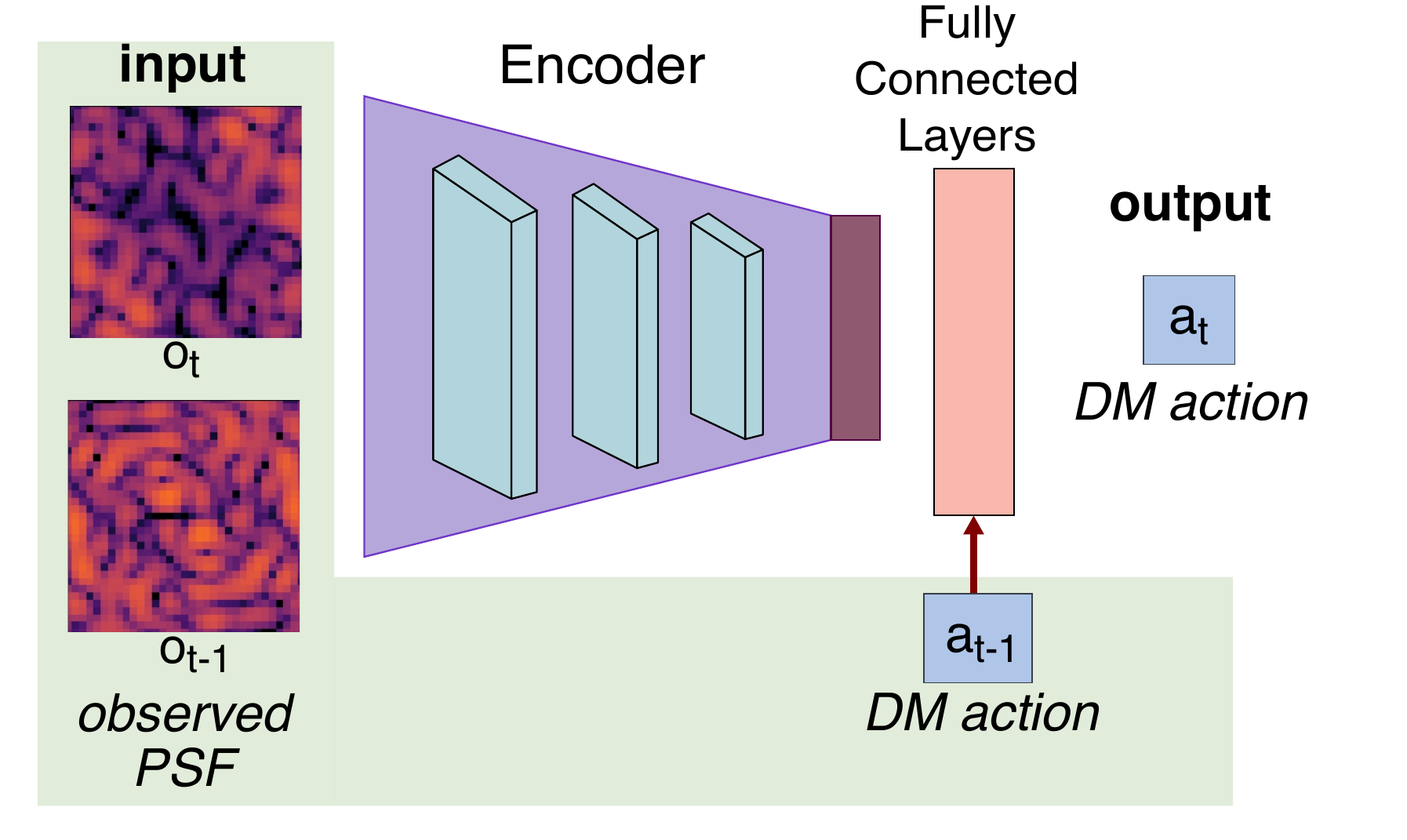}%
\caption{Policy model NN design. In the control loop, the image inputs ($\bm o_t$, and $\bm o_{t-1}$) are focal plane images from the science camera, and while training, the future inputs (Algorithm 1 line 11, $t = 2,3, \cdots H$) are predicted/simulated by the dynamics model.} 
\label{fig:policy}
\end{figure}

\subsection{The policy model}\label{sec:PO4NCPA}
Similarly to the dynamics model, the policy $\pi_\xi(\st)$ also has image and vector inputs. The image inputs are again propagated through convolutional layers, then reshaped into a vector and propagated through some fully connected layers, with the final layer forming the vector output of Zernike coefficients, the action. The previous action (phase diversity action) is concatenated to the first fully connected layer (see Fig.~\ref{fig:policy}). The policy model parameters are then optimized by using the dynamics model. More precisely, we use the dynamics model to generate trajectories of length $H$ (called the planning horizon) starting from states in the data set, with the policy serving as the controller. We then collect rewards and backpropagate through the models to obtain policy gradients and optimize the policy parameters $\xi$. 

Let us define
\begin{equation}\label{eq:reward}
    \hat r_\omega(\st,\action) = - \|\hat {\bm o}_{t+1}\|^2,
\end{equation}
where $\hat {\bm o}_{t+1}$ is obtained from $\tilde {\bm s}_{t+1} = \hat p_\omega(\st, \action)$.
This leads to the approximate policy optimization problem: we approximate Eq. \ref{eq:obj_of_rl}, with learned dynamics and a cropped horizon, that is,
\begin{equation}
    \label{eq:optimization_of_hatr}
    \argmax_\xi \sum_{{\bf s} \in \mathcal{D}}  \sum_{t=1}^{H} \hat r_\omega(\tilde \st, \pi_\xi(\tilde \st)),
\end{equation}
where $H$ is the so-called planning horizon and
\begin{equation*}
    \tilde {\bf s}_1 = {\bf s} \quad \text{and} \quad \tilde {\bf s}_{t+1} = \hat p_\omega(\tilde \st, \pi_\xi(\tilde \st)).
\end{equation*}
To avoid policy optimization from getting stuck in a local minimum, we randomize the horizon length by randomly sampling an integer between $h_{\text{min}}$ and $h_{\text{max}}$. Algorithm \ref{alg:mbpo} presents the complete training procedure for the dynamics model and the policy. The while loop on line 3 cycles through episodes until the performance criteria (i.e., the cumulative reward over episode) converge. The lines 6–16 implement a policy update via policy optimization.

\begin{algorithm}
\caption{Policy Optimization for NCPA}
\label{alg:mbpo}
\begin{algorithmic}[1]
\State Initialize policy and dynamics model parameters $\xi$ and $\omega$ randomly and set ensemble dynamics size $n$
\State Initialize gradient iteration length $K$, batch size $B<|\mathcal{D}|$ and planning horizon $H$ limits
\While{not episode reward converged}
    \State Generate samples $\{s_{t+1}, s_t, a_t \}$  by running policy $\pi_\xi(a_t|s_t)$ for $T$ timesteps (an episode) and append to $\mathcal{D}$
  \State Fit dynamics by minimizing Eq. \eqref{eq:relative_loss} w.r.t $\omega$ using Adam
  \For{iteration $k=1$ to $K$}
      \State Sample a mini batch of $B < |\mathcal{D}|$ states $\{ s_\tau \}$ from $\mathcal{D}$
      \For{each ${\bf s}_\tau$ in the mini batch}
        \State Set $\tilde {\bf s}_1^\tau = {\bf s}_\tau$
        \State Draw $H$ from $(h_{\text{min}},h_{\text{max}})$
        \For{$t=1$ to $H$}
            \State Predict $\action = \pi_\xi(\st)$
            \State Predict $\stp^1, \stp^2, \cdots, \stp^n  = \hat{p}_\omega(\st, \action)$
            \State Calculate $R_t = \sum_{i=1}^n\hat r_\omega(\st^i, \action)$
            \State Ensemble avg next state $\stp = \frac{1}{n}\sum_{i=1}^n \stp^i$
        \EndFor
      \EndFor
      \State Update $\xi$ by taking a gradient step according to $\nabla_\xi \sum_{t=\tau}^{ \tau + H} R_t $ with Adam.
  \EndFor
\EndWhile
\end{algorithmic}
\end{algorithm}

\section{Numerical experiments}\label{sec:results}

\subsection{Simulation description}\label{sec:setup-d}
This section evaluates the performance of PO4NCPA by numerical simulations. First, we demonstrate the method's performance against both static and dynamic NCPA errors, considering two focal planes: one with a perfect (i.e., ideal) coronagraph (PC) and the other without (i.e., standard imaging, SI). Second, we demonstrate the performance of PO4NCPA on a possible use case for the algorithm: WV seeing control for ELT-METIS equipped with a VVC. The examples presented here have been chosen to demonstrate the method's versatility and performance, which is agnostic to imaging conditions (standard imaging or any coronagraphy) and fast enough to cope with dynamic NCPA, while keeping the results section compact.

We used the HCIPy \citep{por2018high} package to simulate telescopes, NCPA aberrations, the deformable mirror, and coronagraphs (PC and VVC).  For the first demonstration (PC or SI), we simulated a 39.3-meter telescope with a circular pupil and no central obstruction, while for the VVC demonstration, we used the latest ELT pupil description released by ESO, which includes one thicker spider arm (386~mm instead of 202~mm) due to the presence of the ELT primary mirror crane (see Fig.~\ref{fig:obs}). We note that this thicker spider effectively breaks the symmetry of the input pupil, which is expected to give this simulation setup an edge in lifting the ambiguity for even Zernike modes. The DM was modeled as a device that applies a command vector comprising Zernike coefficients. The experiments were run with 55 modes (choice discussed in Sec. \ref{sec:discussion}). The focal plane is sampled with three pixels per full-width-at-half-maximum (FWHM), and the image was cropped to contain 5.5 $\lambda/D$. These parameters yielded a $33 \times 33$ pixel focal-plane image that well covers the control radius of the DM (55 Zernike modes). We assumed a frame rate of 10~Hz for the correction of dynamic NCPA (as recommended for WV seeing by \citealp{de2024alf}), and no AO residual WFE.

To keep the results section clear and enable easy comparison between PO4NCPA's static and dynamic NCPA performance, we use the same spatial NCPA spectrum across all experiments. We use the WV seeing spectrum derived by \cite{de2024alf}--the NCPA phase error screens are drawn from the Kolmogorov spectrum with $r_0$ of 95 meters at 11 \textmu m and an outer scale of 500 meters. The resulting median RMS of these phase screens is 286~nm. For the dynamics NCPA cases, we propagate a single layer according to Taylor's frozen flow hypothesis at wind speed of 10~m/s (corresponding to a coherence time of 2.9 sec). We subtract the piston and tip-tilt modes from all of the phase screens; that is,  we assume that the considered instruments have ways to deal with PSF centering, for example, a reflective Lyot stop (with high-speed control-loop, \citealp{singh2014lyot}) or QACITS \citep{Huby+2015} for coronagraphic systems, and normal centroiding for SI. Again, for clarity, all experiments are run at a wavelength of 11 \textmu m (N-band). We note here that the median RMS error (286~nm at N-Band, no piston and tip-tilt) corresponds to 42~nm at 1.65 \textmu m (H-band) in the resulting PSFs. Aberration strength relates to typical NCPAs encountered in NIR/visible HCI instruments like SPHERE \citep[around 50~nm,][]{vigan2019calibration} and MagAO-X  \citep[less than 30~nm,][]{van2021characterizing}.

\begin{table}[t]
    \centering
    \caption{ Simulations parameters}
    \label{table:simulator_parameters}
\begin{tabular}{ l l l} 
 \hline\hline
 \multicolumn{3}{c}{Telescope} \\
 \hline
         Parameter  & Value  &  Units  \\
 \hline
 Telescope diameter  &  39.3  & m     \\
 Sampling frequency   &  10  & Hz        \\
  DM influence functions    &  ``Zernike''  & -  \\
 Number of modes     &  55   & -         \\
 Wavelength  & 11 & \textmu  m \\
 Star flux ($\textrm{mag}_N=1$) &   $3.67 \times 10^9$
   &   \#/frame          \\ 
 Background noise & $1.00 \times 10^8$  & \#/pix/frame  \\
 \hline
 \multicolumn{3}{c}{Water vapor seeing parameters} \\
 \hline
 Fried parameter     &  95  & m @ 11~\textmu  m    \\
 Wind speed   &  10  & m/s    \\
 $L_0$ ($m$)      & 500  &   m        \\ 
 Median RMS wavefront error   &  286  & nm  \\
  \hline
 \multicolumn{3}{c}{PO4NCPA parameters} \\
 \hline
 Horizon limits $(h_{\text{min}},h_{\text{max}})$     &  (2,7)  & steps     \\
 CNN ensemble size        & 5   &   -         \\   
 Dynamics iterations / episode & 8 & steps\\
 Policy iterations / episode & 5 & steps \\
 Training minibatch size  & 64 &  - \\
  \hline 
\end{tabular}

\end{table}

PO4NCPA was tested in the following five experiments representing different environments, using the exact same hyperparameters (see PO4NCPA parameters in Table \ref{table:simulator_parameters}):
\paragraph{Circular pupil + standard imaging (SI)}
    \begin{enumerate}
        \item Static NCPA: We drew a static NCPA error and let the algorithm optimize the focal-plane image, that is, maximize the reward defined in eq. \eqref{eq:reward} through controlling the DM.

        \item Dynamic NCPA: We simulated an evolving NCPA pattern, resembling the expected WV seeing for the ELT/METIS instrument \citep{absil2022impact}. Again, the algorithm aims to maximize the reward by controlling the DM.
    \end{enumerate}
\paragraph{Circular pupil + perfect coronagraph (PC)}
    \begin{enumerate}
        \setcounter{enumi}{2}

        \item Static NCPA: Same as item 1 above.

        \item Dynamic NCPA: Same as item 2 above.
    \end{enumerate}
\paragraph{ELT pupil + VVC + noise}
    \begin{enumerate}
        \setcounter{enumi}{4}

        \item Dynamic NCPA: As a possible use-case, we simulated an ELT/METIS-like system with ELT aperture, dynamic NCPA (WV seeing), photon and background noise, and VVC.
    \end{enumerate}
We compare PO4NCPA against three references: 
\begin{enumerate}
    \item Fitting error: perfect phase correction, that is, the NCPA phase error projected onto the DM Zernike modes, leaving only the high-order fitting errors.
    \item Fitting error + delay error: for the dynamics NCPA case, we couple the perfect phase correction above with a 1-step delay integrator with a gain of 0.8.
    \item Open loop: no phase correction.
\end{enumerate}
We use two types of performance metrics: PSF-related (Strehl ratio, PSF contrast, residual light after the coronagraph), and pupil-plane wavefront-error-related (residual RMS and modal RMS). PO4NCPA optimizes the PSF-related performance, and the wavefront-related performance is a consequence of PSF optimization. We note that these performance metrics do not always align with one another.

\begin{figure*}[t]
\centering
\subfloat[\label{fig:t1}]
        {\includegraphics[trim={0.cm 0cm 0cm 0cm}, clip, width=0.45\textwidth]{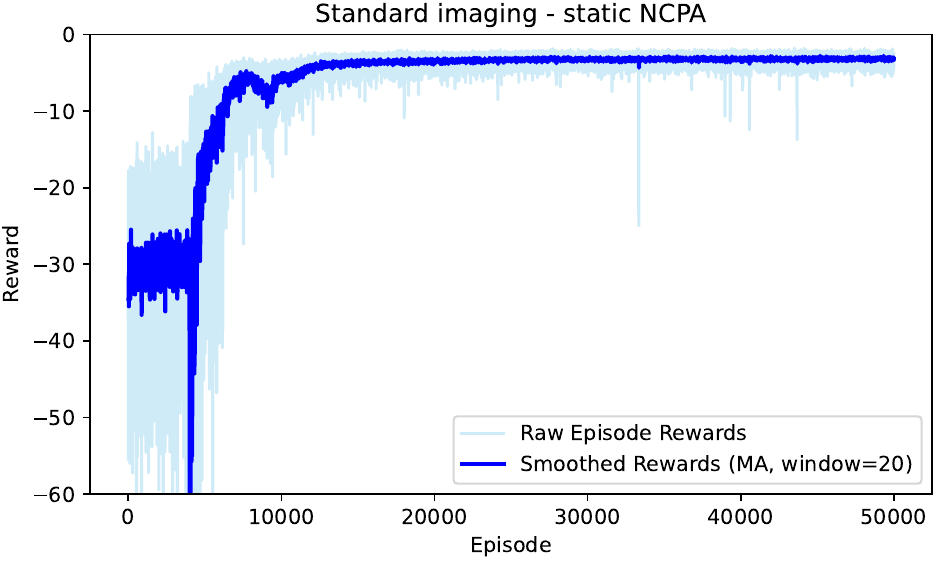}}
    \hfill
\subfloat[\label{fig:t2}]
         {\includegraphics[trim={0.cm 0cm 0cm 0cm}, clip, width=0.45\textwidth]{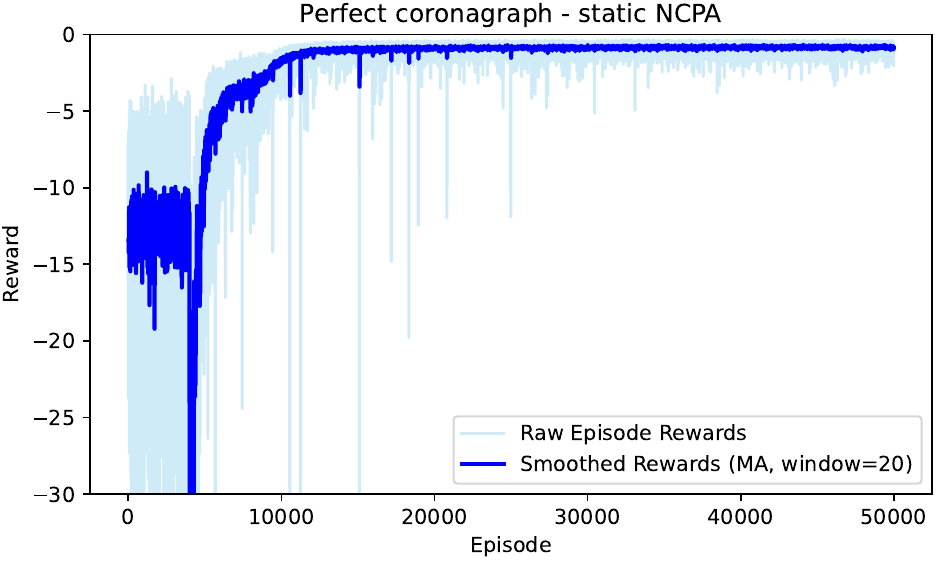}}

\subfloat[ \label{fig:t3}]
         {\includegraphics[trim={0.cm 0cm 0cm 0cm}, clip, width=0.45\textwidth]{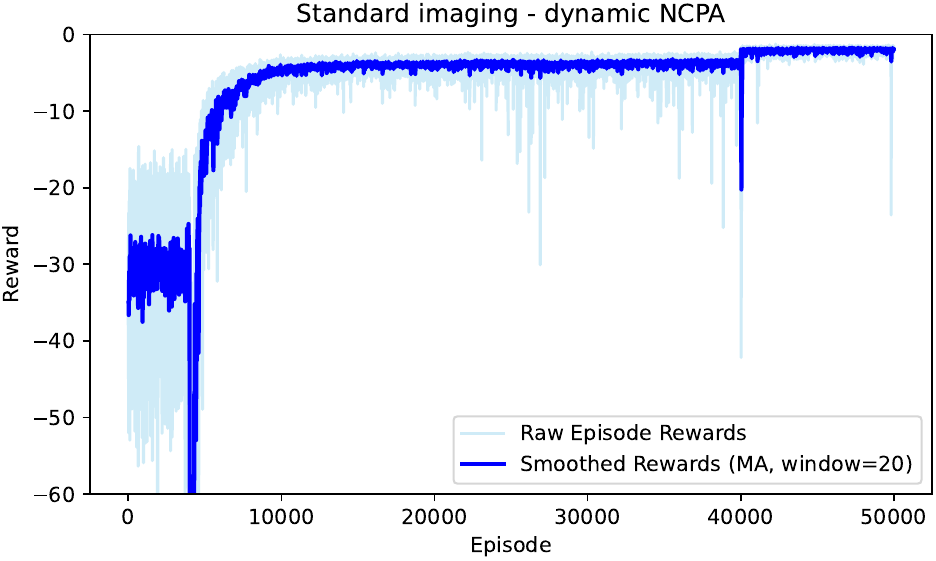}}
    \hfill
\subfloat[ \label{fig:t4}]
        {\includegraphics[trim={0.cm 0cm 0cm 0cm}, clip, width=0.45\textwidth]{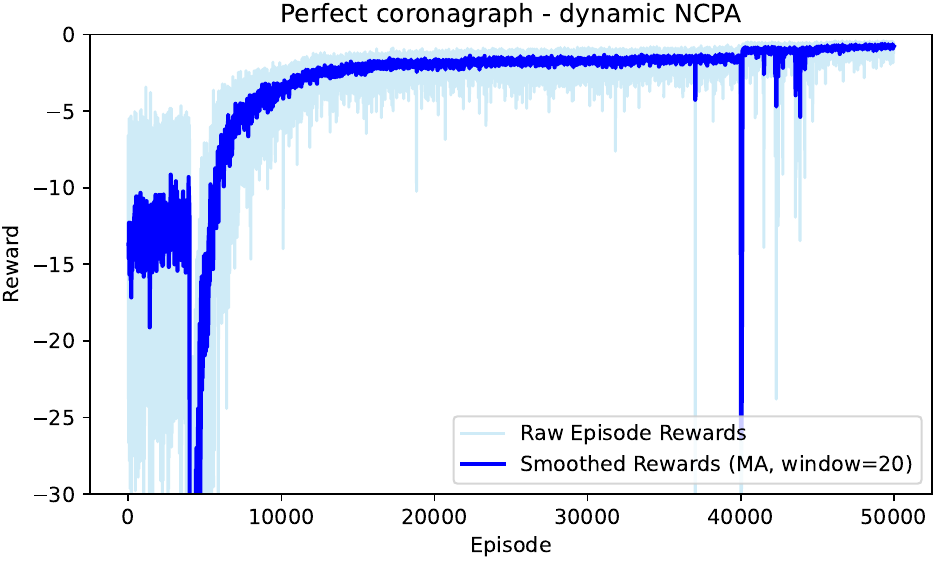}}
      
\caption[Training]
        {Training plots of PO4NCPA on circular pupil with SI and PC. Here we plot the negative cumulative reward (loss) after each episode in the training circle. The light blue curve shows the raw negative reward, and the dark blue shows the smoothed value (moving average). For the dynamic case (c, d), the DM is flattened only at the start of the episode for the first 40k episodes to prevent saturation. Afterward, each episode starts from the previous endpoint, thereby mimicking a continuously updated closed-loop control system. Hence, the bigger episode reward after that.}
    \label{fig:training}
\end{figure*}

\subsection{Circular pupil with SI and PC results}

\subsubsection{Training}
The dynamic and static NCPA codes are trained with the same strategy. The episode length was set to 20 steps. The first 4000 are run with random commands (drawn from a Gaussian distribution with a standard deviation of 0.2 times the total DM response). This phase is called the warm-up. 

For static NCPA, a new phase screen is generated at the beginning of each episode, and the DM is flattened. In the dynamic NCPA cases, the phase screen is propagated at every time step, and the DM is flattened only during the first 40k episodes to avoid starting episodes at a saturated DM position. After the first 40k episodes, each subsequent episode begins from the end position of the previous one, simulating a closed-loop control system in which the new model is continuously updated. 

After each episode, excluding the warm-up, the dynamics are trained for eight gradient steps, and the policy for five gradient steps. The training procedure was continued until the performance (cumulative episode reward) converged. We examined the convergence speed of the method by plotting the cumulative reward (a measure of the distance from the perfect PSF) at the end of each episode (see Fig.~\ref{fig:training}). All cases converge by 50k episodes at the latest. The end of the warm-up phase (4k) and the end of resetting the DM after each episode (at 40k in the dynamic cases) are clearly visible in the plots. At 40k in the dynamic cases, we see some instability at first because PO4NCPA encounters new types of starting positions (not only open-loop WV seeing), but after some episodes, we observe improved performance since PO4NCPA starts from smaller residuals. Further, with the given hyperparameters, the training procedure always converged, producing nearly identical training curves (which is not trivial in RL in general).

\subsubsection{Performance on static NCPA}
\begin{figure*}\centering
\subfloat[\label{fig:re_nocoro}]{
  \includegraphics[trim={0cm 0.00cm 0cm 0cm}, clip,width=0.47\textwidth]{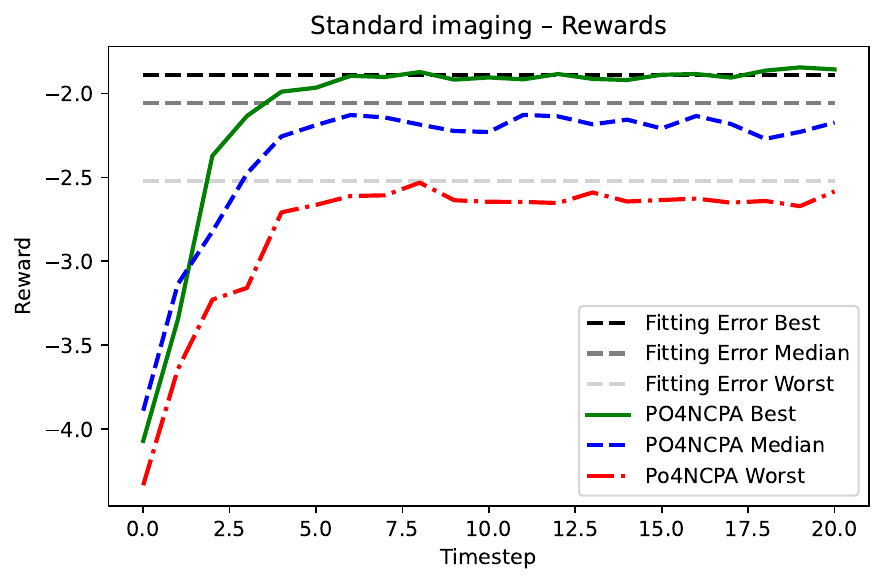}}%
    \hfill
\subfloat[\label{fig:re_coro}]{%
  \includegraphics[trim={0cm 0.00cm 0cm 0cm}, clip,width=0.47\textwidth]{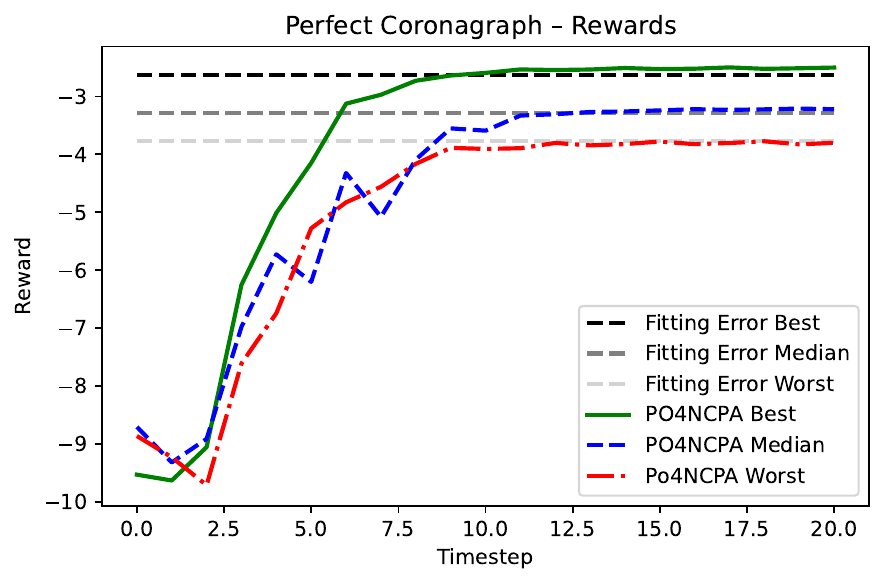}}
  
\subfloat[ \label{fig:rms_nocoro}]{%
  \includegraphics[trim={0cm 0.00cm 0cm 0cm}, clip,width=0.47\textwidth]{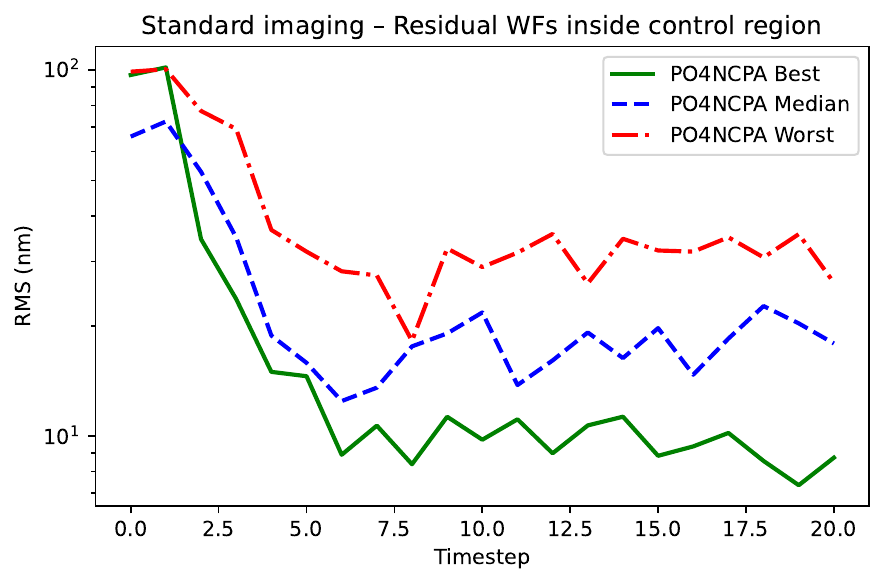}}
    \hfill
\subfloat[ \label{fig:rms_coro}]{%
  \includegraphics[trim={0cm 0.00cm 0cm 0cm}, clip,width=0.47\textwidth]{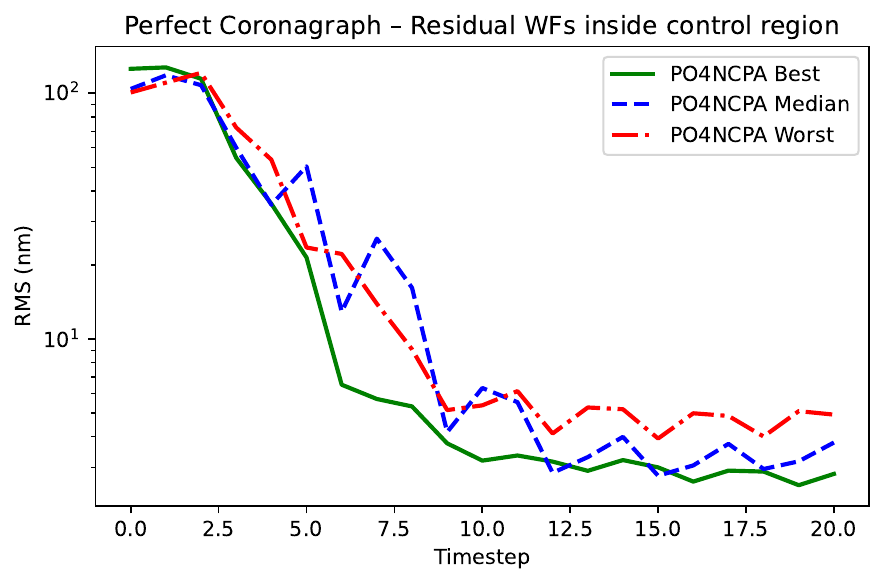}}
      
\caption[Static convergence]
        {PO4NCPA convergence on circular pupil with SI (left) and PC (right) in the case of static NCPA. Top row: reward on time steps. PO4NCPA episodes are shown in green (best), blue (median), red (worst), while black and gray lines correspond to fitting error rewards. Bottom row: PO4NCPA residual wavefront RMS (inside the control region) over the same episodes (fitting error is always zero).}
    \label{fig:convergence}
\end{figure*}

\begin{figure*}
\centering
\subfloat[\label{fig:static_ncpa}]{%
  \includegraphics[trim={0cm 0.00cm 0cm 0cm}, clip,width=0.47\textwidth]{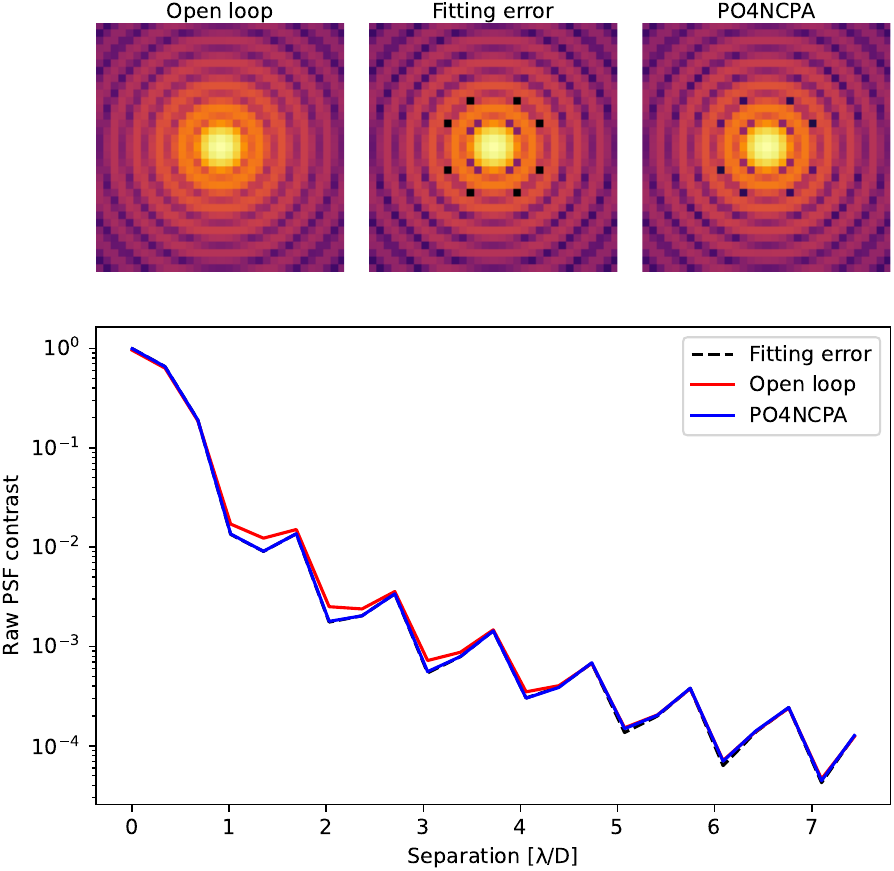}}
  \hfill
\subfloat[\label{fig:static_episode_median_nocoro}]{%
  \includegraphics[trim={0cm 0.00cm 0cm 0cm}, clip,width=0.47\textwidth]{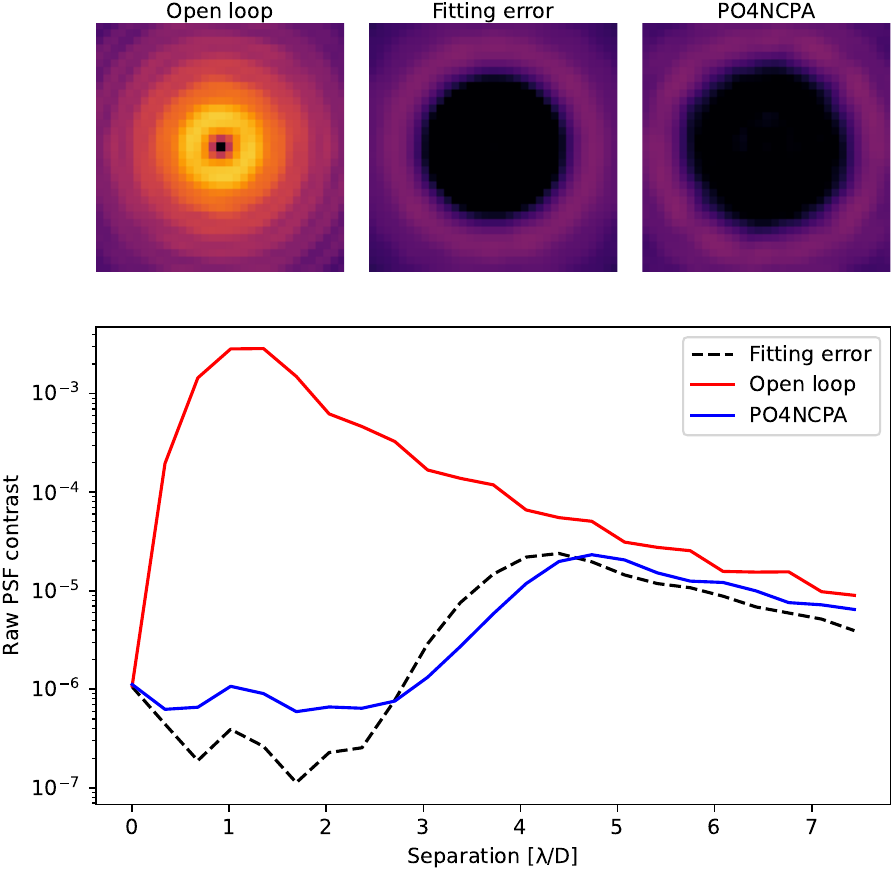}}
  
  \caption{Circular pupil PSF sharpness/raw contrast with static NCPA for (a) standard imaging and (b) perfect coronagraph cases. Here, we take the average (over 500 episodes) of the last PSF frame and plot the resulting PSF and radial average divided by the peak intensity. }
\label{fig:static_PSFs}
\end{figure*}

After the training procedure, we ran 500 episodes with the final policy. Each episode started with a randomly sampled NCPA phase screen and flat DM. We then let the policy run for 20 steps and recorded data at each time step for each episode. The results are visualized in two ways. 

First, we order the episodes by the final reward (Eq. \ref{eq:reward}) at the last step (step \#20). We then plot the temporal evolution of negative reward (correlating to contrast of PSF sharpness) for the best, median, and worst episodes; see Figs.~\ref{fig:re_nocoro} and \ref{fig:re_coro}.
We also plot the corresponding wavefront RMS within the control radius (RMS-fitting RMS) during these episodes in Figs.~\ref{fig:rms_nocoro} and \ref{fig:rms_coro}. We note that the reward is the quantity that the algorithm tries to maximize, and the small RMS is only a consequence of this process. In both PC and SI cases, convergence occurs around 10 timesteps. Without a coronagraph, PO4NCPA achieves a final reward (relative to Strehl) close (but slightly worse) to the reward of fitting error (the colored PO4NCPA lines coincide with black and gray fitting error lines). The corresponding RMS errors are 9~nm for the best, 17~nm for the median, and 26~nm for the worst. In the case of a PC, PO4NCPA achieves equal (worst episode) or better (median and best) reward than the fitting error (Fig.~\ref{fig:re_coro}), meaning that PO4NCPA learns to find a combination of commands that removes more light from the focal plane than the fitting of DM modes -- reward is calculated in the whole focal-plane image, that is, not only inside the dark hole. The corresponding RMS errors are 2~nm (best), 4~nm (median), and 5~nm (worst).

Second, we calculate the mean PSF radial average over the last time step across the 500 different episodes and compare it with the integrator and open-loop (see Fig.~\ref{fig:static_PSFs}).
In the SI case, the PO4NCPA learns to sharpen the slightly distorted PSF (Strehl 95.6\%) to the fitting error limit, yielding an average Strehl ratio of 99.4\%. In contrast, the fitting error Strehl was 99.6\%. In the PC case, PO4NCPA obtains better contrast from 2.7-4.5~$\lambda/D$ than the fitting error, while the fitting error is better at 0-2.7~$\lambda/D$. We note here that PO4NCPA tries to maximize the reward in the whole focal plane (see Sec.~\ref{sec:PO4NCPA}). The total fitting error flux is 0.104\%, and the PO4NCPA is slightly better at 0.102\%, which means that PO4NCPA learns to compromise the contrast near the PSF core to gain contrast farther away by pushing light outside of the field of view (5.5 $\lambda/D$). 

\subsubsection{Dynamic NCPA}

\begin{figure*}
\centering
\subfloat[\label{fig:snipped_nocoro}]{%
  \includegraphics[trim={0cm 0.00cm 0cm 0cm}, clip,width=0.47\textwidth]{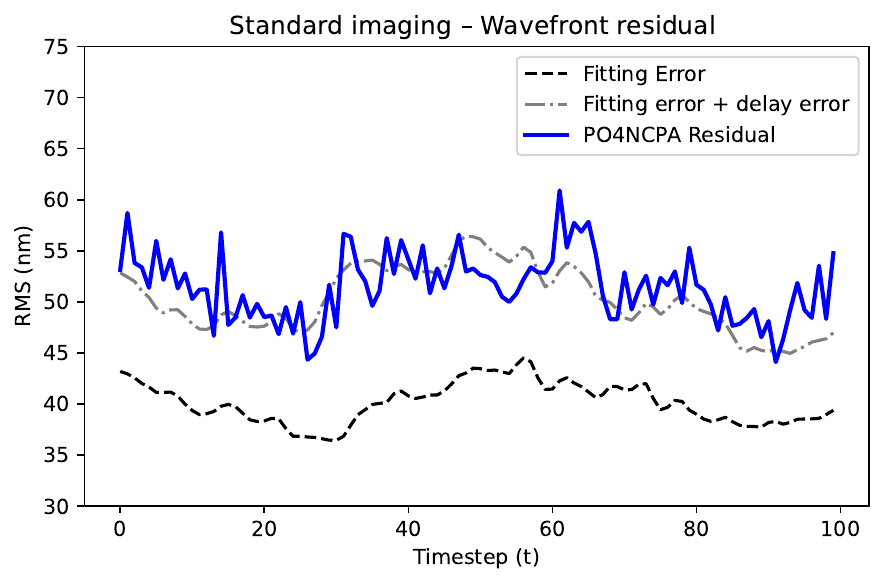}}
  \hfill
\subfloat[\label{fig:snipped}]{%
  \includegraphics[trim={0cm 0.00cm 0cm 0cm}, clip,width=0.47\textwidth]{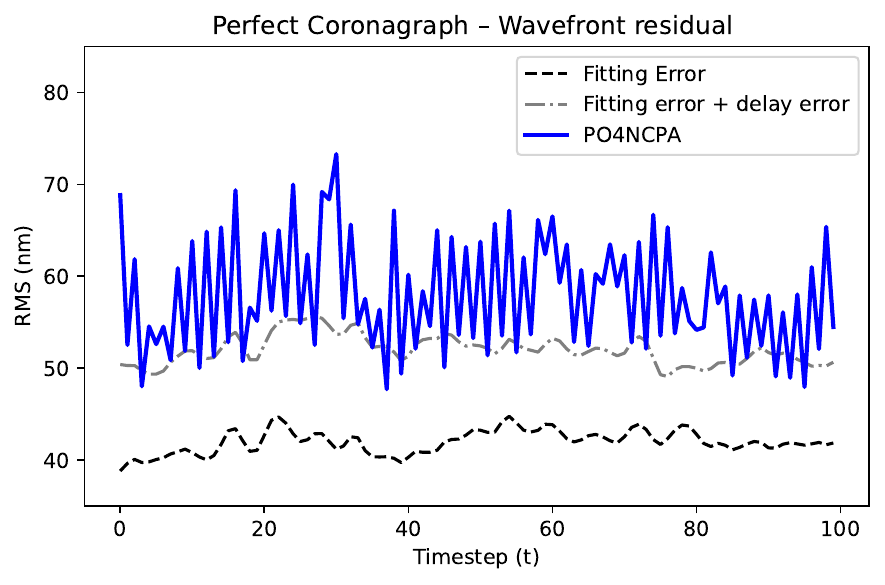}}
  
  \caption{A 100-time-step window during the long exposure dynamic NCPA, showing the wavefront error RMS as a function of time steps. (a) Standard imaging and (b) perfect coronagraph cases. }
\label{fig:dynamic_snipped}
\end{figure*}

\begin{figure*}
\centering
\subfloat[\label{fig:modal_nocoro2}]{%
  \includegraphics[trim={0cm 0.00cm 0cm 0cm}, clip,width=0.47\textwidth]{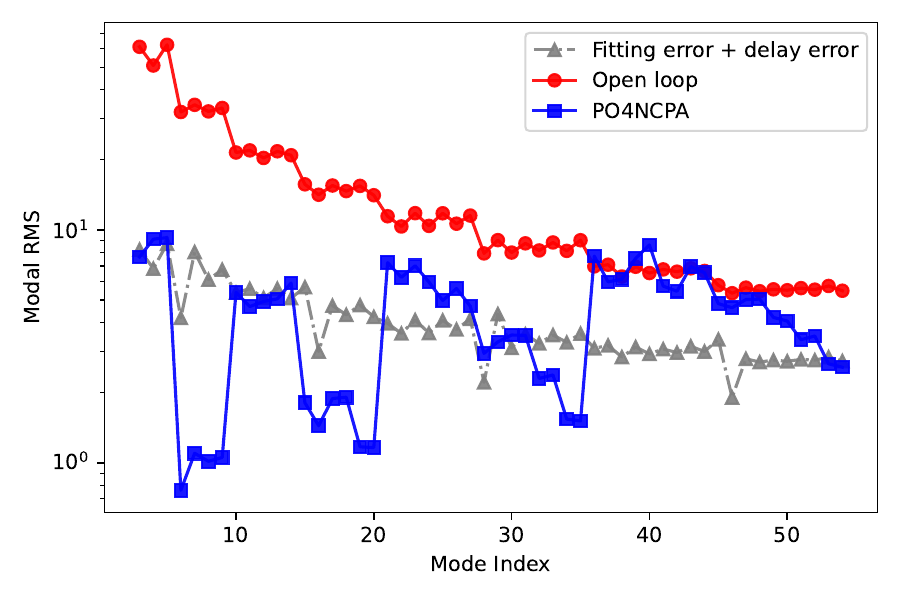}}
  \hfill
\subfloat[\label{fig:modal_coro2}]{%
  \includegraphics[trim={0cm 0.00cm 0cm 0cm}, clip,width=0.47\textwidth]{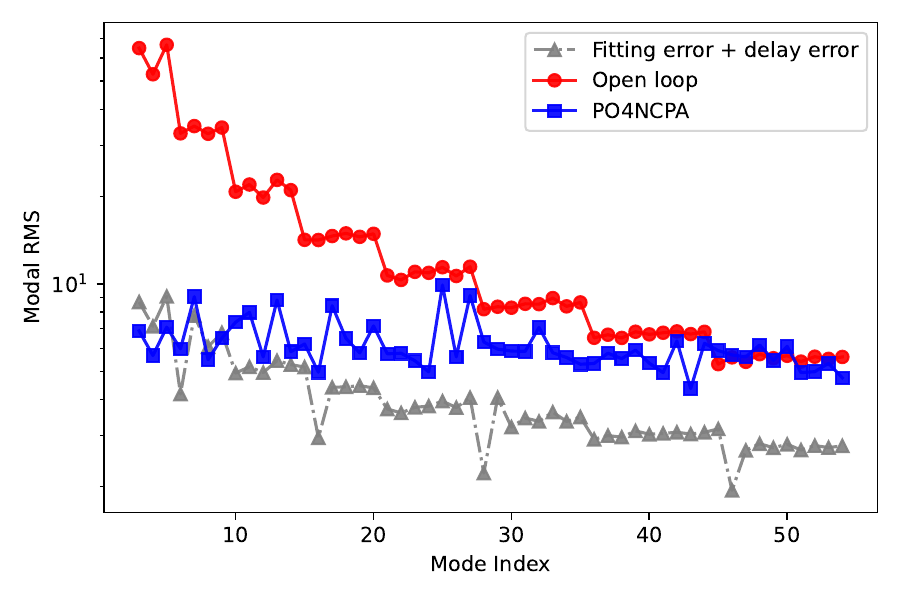}}
  \caption{RMS per mode over the long episode with dynamic NCPA for (a) standard imaging and (b) perfect coronagraph cases.}
\label{fig:modalRMS2}
\end{figure*}

\begin{figure*}
\centering
\subfloat[\label{fig:dynamic}]{%
  \includegraphics[trim={0cm 0.00cm 0cm 0cm}, clip,width=0.47\textwidth]{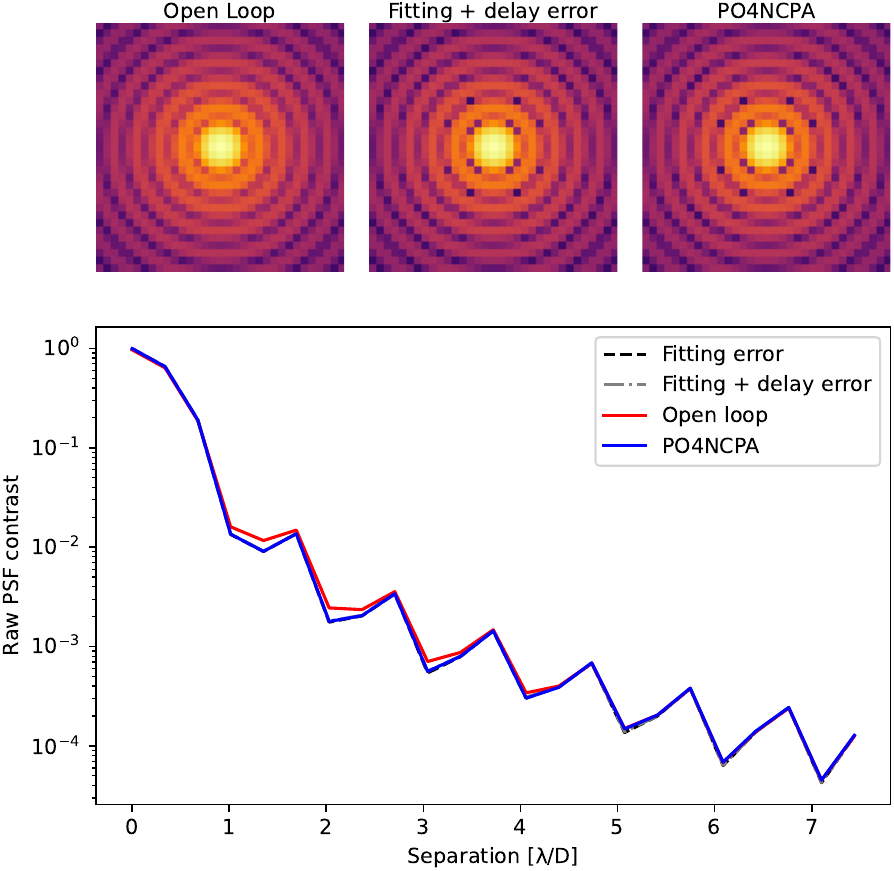}}
  \hfill
\subfloat[\label{fig:dynamics_coro}]{%
  \includegraphics[trim={0cm 0.00cm 0cm 0cm}, clip,width=0.47\textwidth]{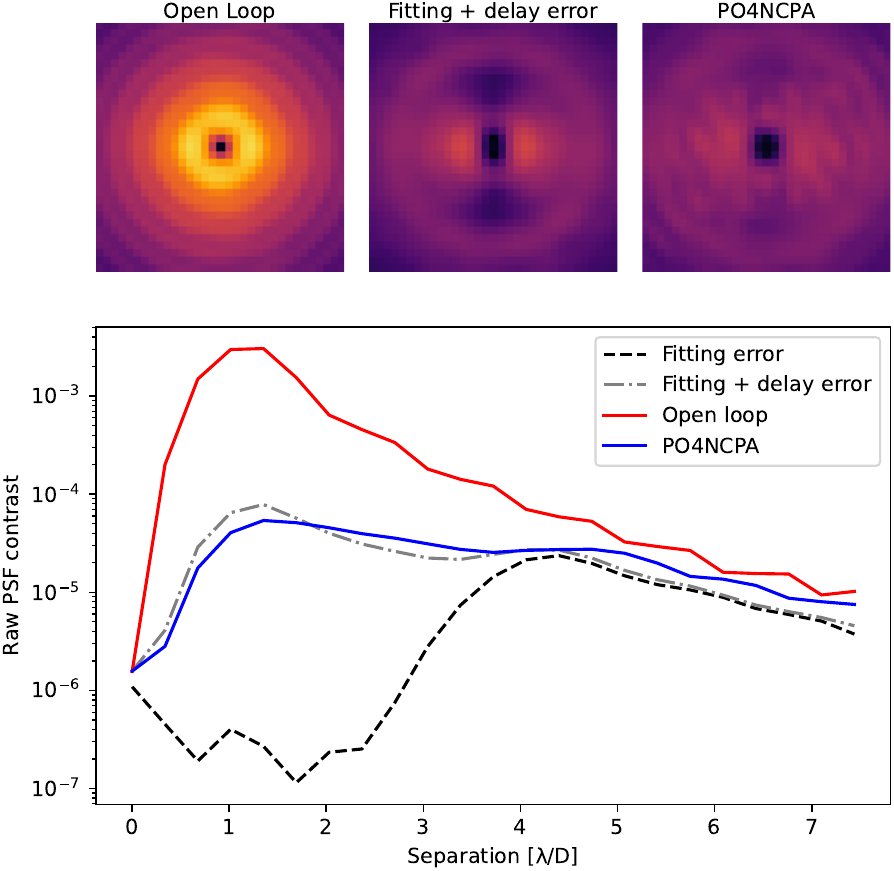}}
  
  \caption{PSF sharpness/contrast with dynamic NCPA for (a) standard imaging and (b) perfect coronagraph cases. }
\label{fig:dynamic_contrast}
\end{figure*}

For dynamic NCPA errors, after the training procedure, we ran the final policy on a single, longer episode. We started with a randomly sampled NCPA phase screen and flat DM, and then let the policy control the DM for 5000 timesteps (i.e, 500 sec) under dynamic NCPA perturbations. We compare PO4NCPA against fitting error, fitting error + time delay error, and the open-loop (no correction).   

Again, the results are visualized in three ways. First, we plot a randomly sampled 100-step sequence of RMS errors (see Fig.~\ref{fig:dynamic_snipped}) for fitting error (dashed black), fitting error + time delay error (gray dot-dashed), and PO4NCPA (blue). Here, the PO4NCPA RMS error follows the ``fitting error + time-delay error'' pattern, suggesting it provides only minimal prediction but can reconstruct the modal coefficients close to the level of perfect reconstruction. 

Second, we plot the modal RMS over the time series to understand how the method behaves with respect to modes; see Fig.~\ref{fig:modalRMS2}. In the SI case, we can clearly see the difference between even modes (sign ambiguity) and odd modes. For odd modes, we observe a factor-of-9 improvement over lower-order modes, suggesting that PO4NCPA learns to predict them. For even modes, performance is diminished, but PO4NCPA still achieves a level of time-delay error on low-order modes, whereas the high-order modes are only slightly corrected. With the PC, there is no difference between even and odd modes, as expected. Instead, we observe that PO4NCPA residuals are smaller only for lower-order modes (1-7), suggesting predictive capability on these modes. We also note that the modal RMS is not the quantity that PO4NCPA tries to minimize; this behavior arises only as a consequence of minimizing the reward function (PSF sharpness/contrast).

Third, Fig. \ref{fig:dynamic_contrast} plots the long-exposure PSFs (over the entire time sequence, excluding the first 20 steps) and their radial averages. For SI, the PO4NCPA PSF closely follows the ``fitting-error'' PSF, with no notable difference. The achieved Strehl ratio is respectively of 95.4\% (open loop), 99.6\% (fitting error + delay error), and 99.4\% (PO4NCPA). The PO4NCPA raw contrast curve closely follows the ``fitting error + delay error'' curve. We also observe that PO4NCPA obtains slightly better contrast at 0-1.8~$\lambda/D$, while the ``fitting error + delay error'' is otherwise slightly better. The total amount of light on the focal plane is respectively 2.75\% (open loop), 0.104\% (fitting error), 0.257\% (fitting error + delay error), and 0.275\% (PO4NCPA).

\subsubsection{Robustness to larger wavefront errors}
The NCPA errors (from the WV seeing spectrum) simulated in the previous subsections are rather small and have a small impact on the Strehl ratio. A natural follow-up question is: can PO4NCPA handle larger wavefront errors? This way it could be used, for example, to dynamically correct the low-wind effect or even drive smaller AO systems without WFS. To this end, we ran an experiment using SI + dynamic NCPA, with the turbulence strength modified by tuning the Fried parameter. We start from the previously defined value of 95~m at N-band and lower it until PO4NCPA no longer performs with the same training procedure and hyperparameter set as before. This time, we do not remove the tip-and-tilt modes; hence, no external centering is needed. Each seeing case (i.e., different $r_0$) is trained separately from the beginning. We run the training procedure and collect the long exposure (5000 frames) Strehl ratio for five different $r_0$ values: 95~m, 30~m, 20~m, 15~m, and 10~m. These values resulted in median piston-removed RMS values of 520~nm, 1336~nm, 1902~nm, 2412~nm, and 3377~nm. The latter corresponding to almost 2 rad RMS for 11µm

\begin{figure}
  \centering
   \includegraphics[trim={0cm 0.00cm 0cm 0cm}, clip, width=0.49\textwidth]{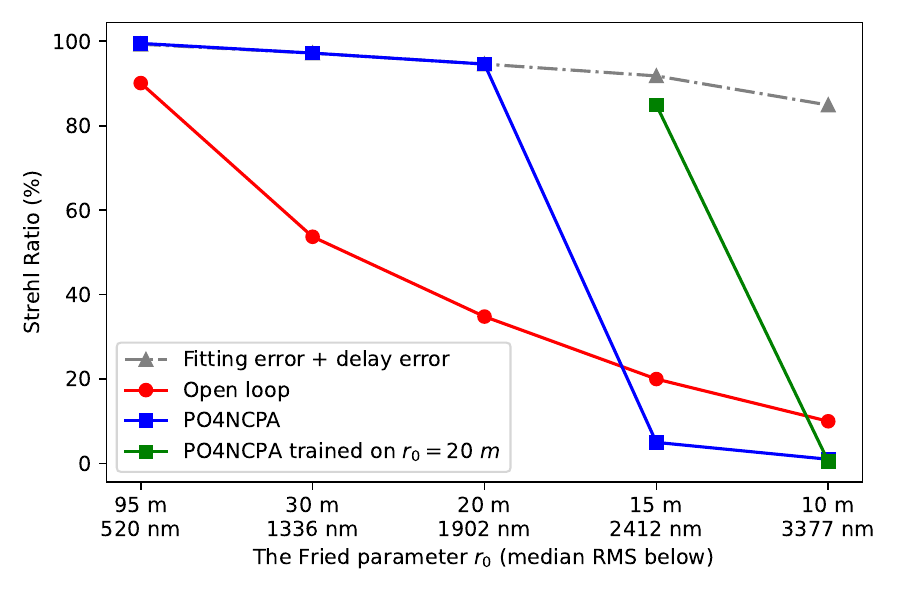}
    \caption{PO4NCPA performance compared against ``Fitting error + delay error'' on larger dynamic wavefront errors in the standard imaging case. The gray dashed line is the ``Fitting error + delay error'', the red line is the open-loop/no-correction Strehl, the blue line is the PO4NCPA trained on the given $r_0$, and the green line is the PO4NCPA trained on $r_0=20$~m applied to $r_0 = 15$~m and 10~m. }
\label{fig:larger_strehl}
\end{figure}

We compare PO4NCPA performance across different wavefront error levels with the previously introduced ``fitting error + delay error'' and open-loop (i.e., no correction). The training procedure converged for Fried parameters ($r_0$) of 95~m, 30~m, and 20~m, and the long exposure Strehl ratio of the final policy follows the ``Fitting error + delay error'' closely (see Fig.~\ref{fig:larger_strehl}). The PO4NCPA training did not converge for $r_0 = 15$~m and $r_0 = 10$~m; instead, it got stuck at an undesired solution (presumably a local minimum of the policy training objective), resulting in poor final performance. However, we also ran the long exposure with controller (i.e., policy) trained on $r_0 = 20$~m with $r_0 = 15$~m and $r_0 = 10$~m: this version of PO4NCPA managed to control the $r_0 = 15$~m case decently (Strehl 85\%, see the green line in Fig.~\ref{fig:larger_strehl}), and also managed to control wavefront errors sometimes for $r_0 = 10$~m, but in a too unstable way to obtain improved long exposure Strehl ratio.

\subsection{ELT pupil with VVC and noise results}
As a possible use-case demonstration, we simulated an ELT/METIS-like system with dynamic NCPA. Compared to the other dynamic NCPA cases, the METIS case uses the ELT pupil, the VVC instead of a PC, and adds photon and strong background noise, typical for ground-based N-band observations (see Table \ref{table:simulator_parameters}).

\begin{figure}
  \centering
   \includegraphics[trim={0cm 0.00cm 0cm 0cm}, clip,width=0.47\textwidth]{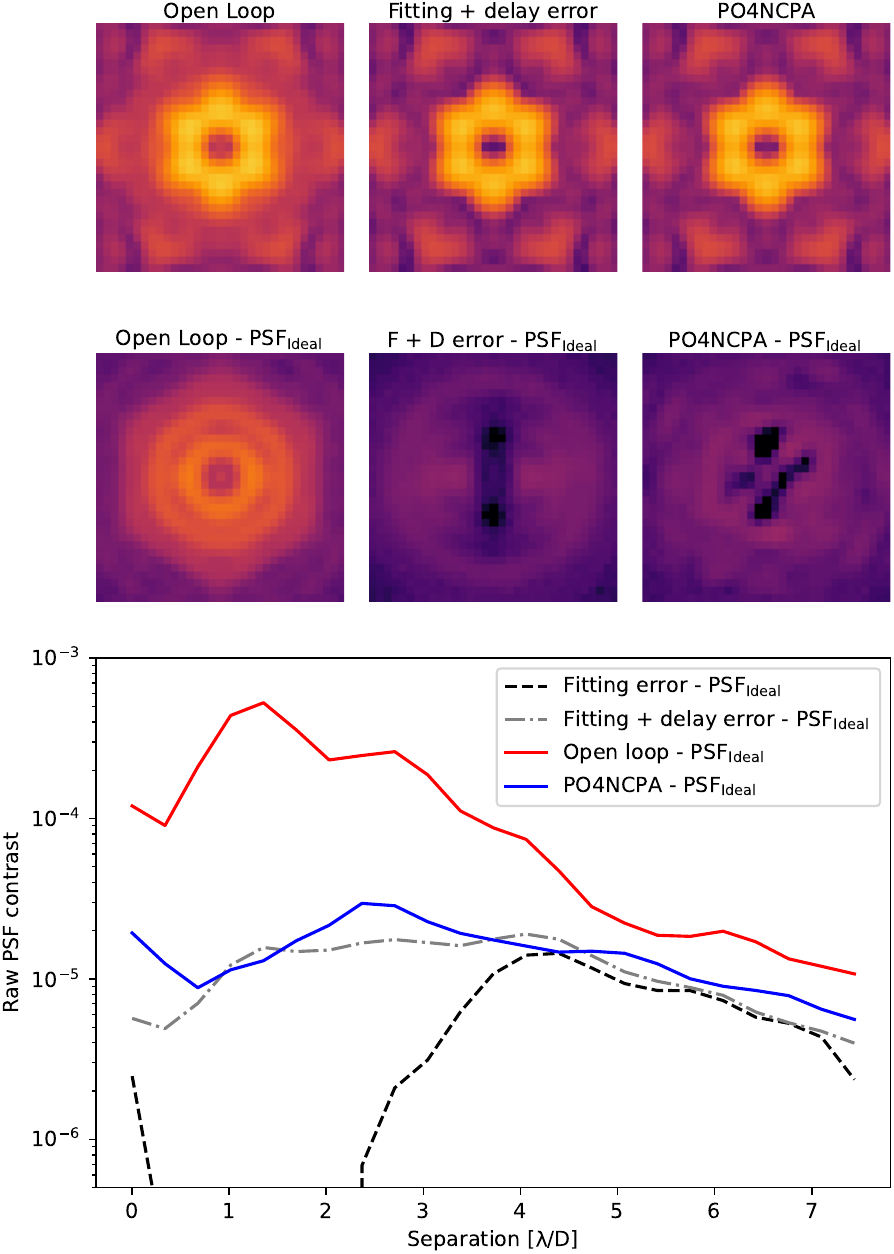}
    \caption{Raw ``residual'' PSF contrast for ELT-METIS. Upper row: recorded PSF. Middle row: recorded PSFs with diffraction patterns (ideal PSF) subtracted. Lower row: the azimuthal average of the subtracted PSFs (i.e., ``Raw residual PSF contras'').}
\label{fig:METIS_contrast}
\end{figure}

For the METIS WV seeing control, we ran the same training procedure, followed by a single 5000-step episode with WV seeing (dynamic NCPA). We only recorded the long-exposure PSF (after training) and calculated the contrast (see Fig.~\ref{fig:METIS_contrast}). We subtract the diffraction pattern (ideal PSF) from the PSFs for better comparison. We again compare the PO4NCPA against fitting error, fitting error + time-delay error, and open-loop. Compared to open-loop, PO4NCPA yields an improvement of up to a factor of 40 at 1.5 $\lambda/D$, and around a factor of 10 at 1.6-3 $\lambda/D$. Similar to the PC experiment, PO4NCPA delivers contrast close to the fitting error + delay error, the latter yielding slightly better overall performance.

\section{Conclusion and future work}\label{sec:discussion}
We introduced a new focal-plane wavefront control algorithm, PO4NCPA. We showed that, when implemented well, it provides near-optimal control, as measured by the reward function (related to the Strehl ratio or post-coronagraphic residual flux), across various NCPA error scenarios. The method makes minimal assumptions about the optical design and can therefore be applied to different optical paths without altering the algorithm. The training phase of PO4NCPA takes relatively long ($\sim$24h including the time spent on the numerical simulation) on standard off-the-shelf GPUs and computers. However, the trained policy (i.e., controller) model is fast to use (inference time < 1ms).  

The method was tested in a numerical simulation, demonstrating robust, promising performance for both static and dynamic NCPA error compensation. In the static NCPA case, PO4NCPA achieves near-optimal performance (in terms of reward). With a perfect coronagraph, PO4NCPA removes more light from the focal plane than the modal least-squares projection, and without a coronagraph, PO4NCPA achieves nearly perfect Strehl. In the dynamic NCPA case, PO4NCPA reaches the performance of the modal least-squares projection combined with a 1-step delay integrator in terms of reward. In terms of wavefront RMS, the performance is farther from the reference method, especially for high-order modes. Additionally, PO4NCPA proved capable of correcting larger wavefront errors. Under the current training setup, the system remained stable up to a median piston-removed RMS of approximately 1902~nm (FP-WFS at 11~µm, corresponding to an open-loop Strehl of about 30\%). The learned policy could even partially correct beyond this range, suggesting that with an optimized learning strategy, the method could adapt to increasingly severe wavefront errors.

In principle, PO4NCPA can perform predictive control in the dynamic NCPA case. Examination of the modal rejection of dynamic SI in Fig.~\ref{fig:modal_nocoro2} indicates that unambiguous modes are corrected significantly more effectively than with the ``fitting error + time-delay error''. In contrast, performance in ambiguous modes is comparable to or worse than time delayed correction of ``fitting error + time-delay error''. This suggests that predictive correction is primarily effective for modes that are unambiguous on the focal plane. With a PC, however, we observe comparable behavior across all modes, since all modes are (to some extent) ambiguous (Fig. \ref{fig:modal_coro2}). We attribute this difference to the requirement for phase diversity in the ambiguous modes. In dynamic NCPA cases, there is always a component of new errors (even with perfect prediction, e.g., \citealp{nousiainen2024power}) for which PO4NCPA must first determine the correct phase sign before applying a correction, which delays the correction process. Consistent with this interpretation, most residual errors are observed along the direction of the wind (see Fig.~\ref{fig:dynamics_coro}). These results suggest the algorithm suits wavefront-sensorless AO (with or without a coronagraph), where correction speed is set by the focal-plane camera's frame rate. However, with the current setup, adding more DM modes did not improve performance — PO4NCPA remains limited to relatively low-order corrections.

Our results show that PO4NCPA also works with the ELT pupil, a vortex coronagraph, and photon and background noise without any modifications to the algorithm. A more thorough analysis of the error terms in these cases is left for future work. As a caveat, the algorithm—like most deep reinforcement learning methods—is somewhat sensitive to hyperparameter choices (e.g., network depth, learning rate, and related settings). Furthermore, deep learning-based control strategies are difficult to analyze theoretically, and formal stability guarantees cannot currently be established.

An obvious line of future work is to study further the METIS VVC WV seeing case. For example, we can perform more accurate numerical simulations that include AO residuals, LWE, intensity variations, and static NCPAs. We will also explore how pupil asymmetries \citep{de2024alf} can help improve the control of even modes. Another interesting direction of future work is the design of a reward function for coronagraphic AO. Here, the reward was defined in the whole focal plane. We could consider defining a smaller one-sided region to dig a deeper dark hole within the area of interest, as well as removing the diffraction pattern subtraction (see Fig.~\ref{fig:obs}) to allow PO4NCPA to apodize beyond the diffraction pattern. 

Further, moving from numerical simulations to lab and on-sky testing requires additional consideration and research. Most prominently, given the method's long training time, it is not feasible to train it on-sky; therefore, a suitable training strategy needs to be studied. For example, first pre-training on a numerical simulation, then fine-tuning with the actual system during the daytime using an internal light source, and finally online training in parallel with closed-loop operation, enabling online adaptation to changing conditions.

\bibliographystyle{aa}
\bibliography{aanda}

@INPROCEEDINGS{Absil+2016,
       author = {{Absil}, Olivier and {Mawet}, Dimitri and {Karlsson}, Mikael and {Carlomagno}, Brunella and {Christiaens}, Valentin and {Defr{\`e}re}, Denis and {Delacroix}, Christian and {Femen{\'\i}a Castella}, Bruno and {Forsberg}, Pontus and {Girard}, Julien and {G{\'o}mez Gonz{\'a}lez}, Carlos A. and {Habraken}, Serge and {Hinz}, Philip M. and {Huby}, Elsa and {Jolivet}, A{\"\i}ssa and {Matthews}, Keith and {Milli}, Julien and {Orban de Xivry}, Gilles and {Pantin}, Eric and {Piron}, Pierre and {Reggiani}, Maddalena and {Ruane}, Garreth J. and {Serabyn}, Gene and {Surdej}, Jean and {Tristram}, Konrad R.~W. and {Vargas Catal{\'a}n}, Ernesto and {Wertz}, Olivier and {Wizinowich}, Peter},
        title = "{Three years of harvest with the vector vortex coronagraph in the thermal infrared}",
    booktitle = {Proc. SPIE Conf.},
         year = 2016,
       volume = {9908},
        month = aug,
          eid = {99080Q},
        pages = {99080Q},
          doi = {10.1117/12.2233289},
archivePrefix = {arXiv},
       eprint = {1607.05003},
 primaryClass = {astro-ph.IM},
       adsurl = {https://ui.adsabs.harvard.edu/abs/2016SPIE.9908E..0QA}
}

@article{terreri2022neural,
  title={Neural networks and PCA coefficients to identify and correct aberrations in adaptive optics},
  author={Terreri, Alessandro and Pedichini, F and Del Moro, D and Causi, G Li and Stangalini, M and Mattioli, M and Piazzesi, R},
  journal={A\&A},
  volume={666},
  pages={A70},
  year={2022},
  publisher={EDP Sciences}
}

@inproceedings{kuznetsov2023prediction,
  title={Prediction of AO corrected PSF for SPHERE/AOF NFM.},
  author={Kuznetsov, Arseniy and Neichel, Benoit and Oberti, Sylvain and Fusco, Thierry},
  booktitle={AO4ELT},
  year={2023}
}

@article{guyon2005limits,
  title={Limits of adaptive optics for high-contrast imaging},
  author={Guyon, Olivier},
  journal={ApJ},
  volume={629},
  number={1},
  pages={592},
  year={2005},
  publisher={IOP Publishing}
}

@article{pou2024integrating,
  title={Integrating supervised and reinforcement learning for predictive control with an unmodulated pyramid wavefront sensor for adaptive optics},
  author={Pou, Bartomeu and Smith, Jeffrey and Quinones, Eduardo and Martin, Mario and Gratadour, Damien},
  journal={Opt. Express},
  volume={32},
  number={21},
  pages={37011--37035},
  year={2024},
  publisher={Optica Publishing Group}
}

@inproceedings{dinis2024upgrading,
  title={Upgrading SPHERE with the second stage AO system SAXO+: frequency-based data-driven controller for adaptive optics},
  author={Dinis, Isaac and Wildi, Fran{\c{c}}ois and S{\'e}gransan, Damien and Gupta, Vaibhav and Karimi, Alireza and Tallon, Michel and Bosc, Isabelle and Langlois, Maud and Loupias, Magali and Bechet, Cl{\'e}mentine and others},
  booktitle={Proc. SPIE Conf.},
  volume={13097},
  pages={1876--1891},
  year={2024},
  organization={SPIE}
}

@article{landman2024making,
  title={Making the unmodulated pyramid wavefront sensor smart-closed-loop demonstration of neural network wavefront reconstruction with magao-x},
  author={Landman, Rico and Haffert, SY and Males, JR and Close, LM and Foster, WB and Van Gorkom, Kyle and Guyon, Olivier and Hedglen, Alex and Kautz, Maggie and Kueny, JK and others},
  journal={A\&A},
  volume={684},
  pages={A114},
  year={2024},
  publisher={EDP Sciences}
}

@article{nousiainen2024laboratory,
  title={Laboratory experiments of model-based reinforcement learning for adaptive optics control},
  author={Nousiainen, Jalo and Engler, Byron and Kasper, Markus and Rajani, Chang and Helin, Tapio and Heritier, C{\'e}dric T and Quanz, Sascha P and Glauser, Adrian M},
  journal={JATIS},
  volume={10},
  number={1},
  pages={019001--019001},
  year={2024},
  publisher={Society of Photo-Optical Instrumentation Engineers}
}

@article{nousiainen2022toward,
  title={Toward on-sky adaptive optics control using reinforcement learning-model-based policy optimization for adaptive optics},
  author={Nousiainen, Jalo and Rajani, Chang and Kasper, Markus and Helin, Tapio and Haffert, Sebastiaan Y and V{\'e}rinaud, Christophe and Males, Jared R and Van Gorkom, Kyle and Close, Laird M and Long, Joseph D and others},
  journal={A\&A},
  volume={664},
  pages={A71},
  year={2022},
  publisher={EDP sciences}
}

@article{landman2025making,
  title={Making the unmodulated pyramid wavefront sensor smart-II. First on-sky demonstration of extreme adaptive optics with deep learning},
  author={Landman, R and Haffert, SY and Long, JD and Males, JR and Close, LM and Foster, WB and Van Gorkom, K and Guyon, O and Hedglen, AD and Johnson, PT and others},
  journal={A\&A},
  volume={696},
  pages={L1},
  year={2025},
  publisher={EDP Sciences}
}

@article{van2022predictive,
  title={Predictive wavefront control on Keck II adaptive optics bench: on-sky coronagraphic results},
  author={van Kooten, Maaike AM and Jensen-Clem, Rebecca and Cetre, Sylvain and Ragland, Sam and Bond, Charlotte Z and Fowler, J and Wizinowich, Peter},
  journal={JATIS},
  volume={8},
  number={2},
  pages={029006--029006},
  year={2022},
  publisher={Society of Photo-Optical Instrumentation Engineers}
}

@article{10.1117/1.JATIS.4.1.019001,
author = {Jared R. Males and Olivier Guyon},
title = {{Ground-based adaptive optics coronagraphic performance under closed-loop predictive control}},
volume = {4},
journal = {JATIS},
number = {1},
publisher = {SPIE},
pages = {019001},
year = {2018},
doi = {10.1117/1.JATIS.4.1.019001},
URL = {https://doi.org/10.1117/1.JATIS.4.1.019001}
}

@inproceedings{give2011pair,
  title={Pair-wise, deformable mirror, image plane-based diversity electric field estimation for high contrast coronagraphy},
  author={Give'on, Amir and Kern, Brian D and Shaklan, Stuart},
  booktitle={Proc. SPIE Conf.},
  volume={8151},
  pages={376--385},
  year={2011},
  organization={SPIE}
}

@article{ruffio2022non,
  title={Non common path aberrations correction},
  author={Ruffio, Jean-Baptiste and Kasper, Markus},
  journal={arXiv e-prints},
  archivePrefix = "arXiv",
  eprint={2211.00775},
  year={2022}
}

@inproceedings{quesnel2022simulator,
  title={A simulator-based autoencoder for focal plane wavefront sensing},
  author={Quesnel, Maxime and Orban de Xivry, Gilles and Absil, Olivier and Louppe, Gilles},
  booktitle={Proc. SPIE Conf.},
  volume={12185},
  pages={982--990},
  year={2022},
  organization={SPIE}
}

@article{haffert2023implicit,
  title={Implicit electric field conjugation: Data-driven focal plane control},
  author={Haffert, SY and Males, JR and Ahn, K and Van Gorkom, K and Guyon, O and Close, LM and Long, JD and Hedglen, AD and Schatz, L and Kautz, M and others},
  journal={A\&A},
  volume={673},
  pages={A28},
  year={2023},
  publisher={EDP Sciences}
}

@article{van2021characterizing,
  title={Characterizing deformable mirrors for the MagAO-X instrument},
  author={Van Gorkom, Kyle and Males, Jared R and Close, Laird M and Lumbres, Jennifer and Hedglen, Alex and Long, Joseph D and Haffert, Sebastiaan Y and Guyon, Olivier and Kautz, Maggie and Schatz, Lauren and others},
  journal={JATIS},
  volume={7},
  number={3},
  pages={039001--039001},
  year={2021},
  publisher={Society of Photo-Optical Instrumentation Engineers}
}

@article{vigan2019calibration,
  title={Calibration of quasi-static aberrations in exoplanet direct-imaging instruments with a zernike phase-mask sensor-iii. on-sky validation in vlt/sphere},
  author={Vigan, Arthur and N’diaye, Mamadou and Dohlen, Kjetil and Sauvage, J-F and Milli, Julien and Zins, G{\'e}rard and Petit, Cyril and Wahhaj, Zahed and Cantalloube, Faustine and Caillat, Amandine and others},
  journal={A\&A},
  volume={629},
  pages={A11},
  year={2019},
  publisher={EDP Sciences}
}

@inproceedings{give2007broadband,
  title={Broadband wavefront correction algorithm for high-contrast imaging systems},
  author={Give'on, Amir and Kern, Brian and Shaklan, Stuart and Moody, Dwight C and Pueyo, Laurent},
  booktitle={Proc. SPIE Conf.},
  volume={6691},
  pages={63--73},
  year={2007},
  organization={SPIE}
}

@INPROCEEDINGS{2023ASPC..534..799C,
       author = {{Currie}, T. and {Biller}, B. and {Lagrange}, A. and {Marois}, C. and {Guyon}, O. and {Nielsen}, E.~L. and {Bonnefoy}, M. and {De Rosa}, R.~J.},
        title = "{Direct Imaging and Spectroscopy of Extrasolar Planets}",
    booktitle = {Protostars and Planets VII},
         year = 2023,
       editor = {{Inutsuka}, S. and {Aikawa}, Y. and {Muto}, T. and {Tomida}, K. and {Tamura}, M.},
       series = {Astronomical Society of the Pacific Conference Series},
       volume = {534},
        month = jul,
        pages = {799},
          doi = {10.48550/arXiv.2205.05696},
archivePrefix = {arXiv},
       eprint = {2205.05696},
 primaryClass = {astro-ph.EP},
       adsurl = {https://ui.adsabs.harvard.edu/abs/2023ASPC..534..799C}
}

@article{riaud2012instantaneous,
  title={Instantaneous phase retrieval with the vector vortex coronagraph-Theoretical and optical implementation},
  author={Riaud, P and Mawet, D and Magette, A},
  journal={A\&A},
  volume={545},
  pages={A151},
  year={2012},
  publisher={EDP Sciences}
}

@inproceedings{de2024vortex,
  title={Vortex coronagraph: revisiting the phase retrieval properties via Zernike analysis},
  author={Orban de Xivry, G and Absil, Olivier},
  booktitle={Proc. SPIE Conf.},
  volume={13097},
  pages={982--988},
  year={2024},
  organization={SPIE}
}

@inproceedings{absil2022impact,
  title={Impact of water vapor seeing on mid-infrared high-contrast imaging at ELT scale},
  author={Absil, Olivier and Delacroix, Christian and Orban de Xivry, Gilles and Pathak, Prashant and Willson, Matthew and Berio, Philippe and Van Boekel, Roy and Matter, Alexis and Defr{\`e}re, Denis and Burtscher, Leo and others},
  booktitle={Proc. SPIE Conf.},
  volume={12185},
  pages={298--310},
  year={2022},
  organization={SPIE}
}

@article{nousiainen2021adaptive,
  title={Adaptive optics control using model-based reinforcement learning},
  author={Nousiainen, Jalo and Rajani, Chang and Kasper, Markus and Helin, Tapio},
  journal={Opt. Express},
  volume={29},
  number={10},
  pages={15327--15344},
  year={2021},
  publisher={Optical Society of America}
}

@article{ke2019self,
  title={Self-learning control for wavefront sensorless adaptive optics system through deep reinforcement learning},
  author={Ke, Hu and Xu, Bing and Xu, Zhenxing and Wen, Lianghua and Yang, Ping and Wang, Shuai and Dong, Lizhi},
  journal={Optik},
  volume={178},
  pages={785--793},
  year={2019},
  publisher={Elsevier}
}

@inproceedings{nagabandi2018neural,
  title={Neural network dynamics for model-based deep reinforcement learning with model-free fine-tuning},
  author={Nagabandi, Anusha and Kahn, Gregory and Fearing, Ronald S and Levine, Sergey},
  booktitle={2018 IEEE International Conference on Robotics and Automation (ICRA)},
  pages={7559--7566},
  year={2018},
  organization={IEEE}
}

@inproceedings{deisenroth2011pilco,
  title={PILCO: A model-based and data-efficient approach to policy search},
  author={Deisenroth, Marc and Rasmussen, Carl E},
  booktitle={ICML-11},
  pages={465--472},
  year={2011},
  organization={Citeseer}
}

@article{cavarroc2006fundamental,
  title={Fundamental limitations on Earth-like planet detection with extremely large telescopes},
  author={Cavarroc, Celine and Boccaletti, A and Baudoz, P and Fusco, T and Rouan, D},
  journal={A\&A},
  volume={447},
  number={1},
  pages={397--403},
  year={2006},
  publisher={EDP Sciences}
}

@article{Landman:20,
author = {R. Landman and S. Y. Haffert},
journal = {Opt. Express},
number = {11},
pages = {16644--16657},
publisher = {OSA},
title = {Nonlinear wavefront reconstruction with convolutional neural networks for Fourier-based wavefront sensors},
volume = {28},
month = {May},
year = {2020},
url = {http://www.opticsexpress.org/abstract.cfm?URI=oe-28-11-16644},
doi = {10.1364/OE.389465}
}

@article{kingma2014adam,
  title={Adam: A method for stochastic optimization},
  author={Kingma, Diederik P and Ba, Jimmy},
  journal={ArXiv e-prints},
  archivePrefix = "arXiv",
  eprint = {1412.6980},
  year={2014}
}

@article{guyon2017adaptive,
  title={Adaptive optics predictive control with empirical orthogonal functions (EOFs)},
  author={Guyon, Olivier and Males, Jared},
  journal={ArXiv preprint},
  archivePrefix = "arXiv",
  eprint = {1707.00570},
  year={2017}
}

@inproceedings{chua2018deep,
  title={Deep reinforcement learning in a handful of trials using probabilistic dynamics models},
  author={Chua, Kurtland and Calandra, Roberto and McAllister, Rowan and Levine, Sergey},
  booktitle={NeurIPS},
  pages={4754--4765},
  year={2018}
}

@ARTICLE{2004ApJ...615L..61M,
       author = {{Marois}, Christian and {Racine}, Ren{\'e} and {Doyon}, Ren{\'e} and {Lafreni{\`e}re}, David and {Nadeau}, Daniel},
        title = "{Differential Imaging with a Multicolor Detector Assembly: A New Exoplanet Finder Concept}",
      journal = {\apjl},
         year = 2004,
        month = nov,
       volume = {615},
       number = {1},
        pages = {L61-L64},
          doi = {10.1086/426077},
archivePrefix = {arXiv},
       eprint = {astro-ph/0410010},
 primaryClass = {astro-ph},
       adsurl = {https://ui.adsabs.harvard.edu/abs/2004ApJ...615L..61M}
}

@ARTICLE{2006ApJ...641..556M,
       author = {{Marois}, Christian and {Lafreni{\`e}re}, David and {Doyon}, Ren{\'e} and {Macintosh}, Bruce and {Nadeau}, Daniel},
        title = "{Angular Differential Imaging: A Powerful High-Contrast Imaging Technique}",
      journal = {\apj},
         year = 2006,
        month = apr,
       volume = {641},
       number = {1},
        pages = {556-564},
          doi = {10.1086/500401},
archivePrefix = {arXiv},
       eprint = {astro-ph/0512335},
 primaryClass = {astro-ph},
       adsurl = {https://ui.adsabs.harvard.edu/abs/2006ApJ...641..556M}
}

@ARTICLE{2015A&A...576A..59S,
       author = {{Snellen}, I. and {de Kok}, R. and {Birkby}, J.~L. and {Brandl}, B. and {Brogi}, M. and {Keller}, C. and {Kenworthy}, M. and {Schwarz}, H. and {Stuik}, R.},
        title = "{Combining high-dispersion spectroscopy with high contrast imaging: Probing rocky planets around our nearest neighbors}",
      journal = {\aap},
         year = 2015,
        month = apr,
       volume = {576},
          eid = {A59},
        pages = {A59},
          doi = {10.1051/0004-6361/201425018},
archivePrefix = {arXiv},
       eprint = {1503.01136},
 primaryClass = {astro-ph.EP},
       adsurl = {https://ui.adsabs.harvard.edu/abs/2015A&A...576A..59S}
}

@ARTICLE{2021A&A...646A.150O,
       author = {{Otten}, G.~P.~P.~L. and {Vigan}, A. and {Muslimov}, E. and {N'Diaye}, M. and {Choquet}, E. and {Seemann}, U. and {Dohlen}, K. and {Houll{\'e}}, M. and {Cristofari}, P. and {Phillips}, M.~W. and {Charles}, Y. and {Baraffe}, I. and {Beuzit}, J. -L. and {Costille}, A. and {Dorn}, R. and {El Morsy}, M. and {Kasper}, M. and {Lopez}, M. and {Mordasini}, C. and {Pourcelot}, R. and {Reiners}, A. and {Sauvage}, J. -F.},
        title = "{Direct characterization of young giant exoplanets at high spectral resolution by coupling SPHERE and CRIRES+}",
      journal = {\aap},
         year = 2021,
        month = feb,
       volume = {646},
          eid = {A150},
        pages = {A150},
          doi = {10.1051/0004-6361/202038517},
archivePrefix = {arXiv},
       eprint = {2009.01841},
 primaryClass = {astro-ph.IM},
       adsurl = {https://ui.adsabs.harvard.edu/abs/2021A&A...646A.150O}
}

@article{guyon2018extreme,
  title={Extreme adaptive optics},
  author={Guyon, Olivier},
  journal={ARA\&A},
  volume={56},
  pages={315--355},
  year={2018},
  publisher={Annual Reviews}
}

@ARTICLE{1982OptEn..21..829G,
       author = {{Gonsalves}, R.~A.},
        title = "{Phase Retrieval And Diversity In Adaptive Optics}",
      journal = {Optical Engineering},
         year = 1982,
        month = oct,
       volume = {21},
       number = {5},
        pages = {829},
          doi = {10.1117/12.7972989},
       adsurl = {https://ui.adsabs.harvard.edu/abs/1982OptEn..21..829G}
}

@article{jovanovic2015subaru,
  title={The Subaru coronagraphic extreme adaptive optics system: enabling high-contrast imaging on solar-system scales},
  author={Jovanovic, N and Martinache, Frantz and Guyon, Olivier and Clergeon, Christophe and Singh, Garima and Kudo, Tomoyuki and Garrel, Vincent and Newman, Kevin and Doughty, D and Lozi, Julien and others},
  journal={PASP},
  volume={127},
  number={955},
  pages={890},
  year={2015},
  publisher={IOP Publishing}
}

@inproceedings{brandl2024final,
  title={Final design and status of the Mid-Infrared ELT Imager and Spectrograph, METIS},
  author={Brandl, B and Absil, O and Feldt, M and Garcia, P and Glasse, A and Guedel, M and Labadie, L and Meyer, M and Pantin, E and Quanz, S and others},
  booktitle={Proc. SPIE},
  volume={13096},
  year={2024}
}

@inproceedings{macintosh2008gemini,
  title={The Gemini Planet Imager: from science to design to construction},
  author={Macintosh, Bruce A and Graham, James R and Palmer, David W and Doyon, Ren{\'e} and Dunn, Jennifer and Gavel, Donald T and Larkin, James and Oppenheimer, Ben and Saddlemyer, Les and Sivaramakrishnan, Anand and others},
  booktitle={Proc. SPIE Conf},
  volume={7015},
  pages={315--327},
  year={2008},
  organization={SPIE}
}

@inproceedings{milli2018low,
  title={Low wind effect on VLT/SPHERE: impact, mitigation strategy, and results},
  author={Milli, Julien and Kasper, Markus and Bourget, Pierre and Pannetier, Cyril and Mouillet, David and Sauvage, J-F and Reyes, Claudia and Fusco, Thierry and Cantalloube, Faustine and Tristam, Konrad and others},
  booktitle={Proc. SPIE Conf},
  volume={10703},
  pages={752--771},
  year={2018},
  organization={SPIE}
}

@article{martinache2013asymmetric,
  title={The asymmetric pupil Fourier wavefront sensor},
  author={Martinache, Frantz},
  journal={PASP},
  volume={125},
  number={926},
  pages={422},
  year={2013},
  publisher={IOP Publishing}
}

@inproceedings{por2018high,
  title={High Contrast Imaging for Python (HCIPy): an open-source adaptive optics and coronagraph simulator},
  author={Por, Emiel H and Haffert, Sebastiaan Y and Radhakrishnan, Vikram M and Doelman, David S and Van Kooten, Maaike and Bos, Steven P},
  booktitle={Proc. SPIE Conf},
  volume={10703},
  pages={1112--1125},
  year={2018},
  organization={SPIE}
}

@inproceedings{gonsalves2010sequential,
  title={Sequential diversity imaging: phase diversity with AO changes as the diversities},
  author={Gonsalves, Robert A},
  booktitle={Frontiers in Optics},
  pages={FWV1},
  year={2010},
  organization={Optica Publishing Group}
}

@article{nousiainen2024power,
  title={Power of prediction: spatiotemporal Gaussian process modeling for predictive control in slope-based wavefront sensing},
  author={Nousiainen, Jalo and Puska, Juha-Pekka and Helin, Tapio and Hyv{\"o}nen, Nuutti and Kasper, Markus},
  journal={JATIS},
  volume={10},
  number={3},
  pages={039001--039001},
  year={2024},
  publisher={Society of Photo-Optical Instrumentation Engineers}
}

@inproceedings{keller2012extremely,
  title={Extremely fast focal-plane wavefront sensing for extreme adaptive optics},
  author={Keller, Christoph U and Korkiakoski, Visa and Doelman, Niek and Fraanje, Rufus and Andrei, Raluca and Verhaegen, Michel},
  booktitle={Proc. SPIE Conf},
  volume={8447},
  pages={749--758},
  year={2012},
  organization={SPIE}
}

@inproceedings{de2024alf,
  title={ALF: an asymmetric Lyot wavefront sensor for the ELT/METIS vortex coronagraph},
  author={Orban de Xivry, G and Absil, Olivier and Delacroix, Christian and Pathak, Prashant and Quesnel, Maxime and Bertram, Thomas},
  booktitle={Proc. SPIE Conf.},
  volume={13097},
  pages={974--981},
  year={2024},
  organization={SPIE}
}

@article{singh2014lyot,
  title={Lyot-based low order wavefront sensor for phase-mask coronagraphs: principle, simulations and laboratory experiments},
  author={Singh, Garima and Martinache, Frantz and Baudoz, Pierre and Guyon, Olivier and Matsuo, Taro and Jovanovic, Nemanja and Clergeon, Christophe},
  journal={PASP},
  volume={126},
  number={940},
  pages={586},
  year={2014},
  publisher={IOP Publishing}
}

@article{mawet2005annular,
  title={Annular groove phase mask coronagraph},
  author={Mawet, Dimitri and Riaud, Pierre and Absil, Olivier and Surdej, Jean},
  journal={ApJ},
  volume={633},
  number={2},
  pages={1191},
  year={2005},
  publisher={IOP Publishing}
}

@article{guyon2009coronagraphic,
  title={Coronagraphic low-order wave-front sensor: principle and application to a phase-induced amplitude coronagraph},
  author={Guyon, Olivier and Matsuo, Taro and Angel, Roger},
  journal={ApJ},
  volume={693},
  number={1},
  pages={75},
  year={2009},
  publisher={IOP Publishing}
}

@article{durech2021wavefront,
  title={Wavefront sensor-less adaptive optics using deep reinforcement learning},
  author={Durech, Eduard and Newberry, William and Franke, Jonas and Sarunic, Marinko V},
  journal={Biomed. Opt. Express},
  volume={12},
  number={9},
  pages={5423--5438},
  year={2021},
  publisher={OSA}
}

@ARTICLE{1990Angel,
       author = {{Angel}, J.~R.~P. and {Wizinowich}, P. and {Lloyd-Hart}, M. and {Sandler}, D.},
        title = "{Adaptive optics for array telescopes using neural-network techniques}",
      journal = {Nat},
         year = 1990,
        month = nov,
       volume = {348},
       number = {6298},
        pages = {221-224},
          doi = {10.1038/348221a0},
       adsurl = {https://ui.adsabs.harvard.edu/abs/1990Natur.348..221A}
}

@ARTICLE{2023Wong,
       author = {{Wong}, Alison P. and {Norris}, Barnaby R.~M. and {Deo}, Vincent and {Tuthill}, Peter G. and {Scalzo}, Richard and {Sweeney}, David and {Ahn}, Kyohoon and {Lozi}, Julien and {Vievard}, S{\'e}bastien and {Guyon}, Olivier},
        title = "{Nonlinear Wave Front Reconstruction from a Pyramid Sensor using Neural Networks}",
      journal = {\pasp},
         year = 2023,
        month = nov,
       volume = {135},
       number = {1053},
          eid = {114501},
        pages = {114501},
          doi = {10.1088/1538-3873/acfdcb},
archivePrefix = {arXiv},
       eprint = {2311.02595},
 primaryClass = {astro-ph.IM},
       adsurl = {https://ui.adsabs.harvard.edu/abs/2023PASP..135k4501W}
}

@ARTICLE{2022Quesnel,
       author = {{Quesnel}, M. and {Orban de Xivry}, G. and {Louppe}, G. and {Absil}, O.},
        title = "{A deep learning approach for focal-plane wavefront sensing using vortex phase diversity}",
      journal = {\aap},
         year = 2022,
        month = dec,
       volume = {668},
          eid = {A36},
        pages = {A36},
          doi = {10.1051/0004-6361/202143001},
archivePrefix = {arXiv},
       eprint = {2210.00632},
 primaryClass = {astro-ph.IM},
       adsurl = {https://ui.adsabs.harvard.edu/abs/2022A&A...668A..36Q}
}

@ARTICLE{2021Orban,
       author = {{Orban de Xivry}, G. and {Quesnel}, M. and {Vanberg}, P. -O. and {Absil}, O. and {Louppe}, G.},
        title = "{Focal plane wavefront sensing using machine learning: performance of convolutional neural networks compared to fundamental limits}",
      journal = {\mnras},
         year = 2021,
        month = aug,
       volume = {505},
       number = {4},
        pages = {5702-5713},
          doi = {10.1093/mnras/stab1634},
archivePrefix = {arXiv},
       eprint = {2106.04456},
 primaryClass = {astro-ph.IM},
       adsurl = {https://ui.adsabs.harvard.edu/abs/2021MNRAS.505.5702O}
}

@ARTICLE{2022Skaf,
       author = {{Skaf}, Nour and {Guyon}, Olivier and {Gendron}, {\'E}ric and {Ahn}, Kyohoon and {Bertrou-Cantou}, Arielle and {Boccaletti}, Anthony and {Cranney}, Jesse and {Currie}, Thayne and {Deo}, Vincent and {Edwards}, Billy and {Ferreira}, Florian and {Gratadour}, Damien and {Lozi}, Julien and {Norris}, Barnaby and {Sevin}, Arnaud and {Vidal}, Fabrice and {Vievard}, S{\'e}bastien},
        title = "{On-sky validation of image-based adaptive optics wavefront sensor referencing}",
      journal = {\aap},
         year = 2022,
        month = mar,
       volume = {659},
          eid = {A170},
        pages = {A170},
          doi = {10.1051/0004-6361/202141514},
archivePrefix = {arXiv},
       eprint = {2110.14997},
 primaryClass = {astro-ph.IM},
       adsurl = {https://ui.adsabs.harvard.edu/abs/2022A&A...659A.170S}
}

@ARTICLE{2020Bos,
       author = {{Bos}, S.~P. and {Vievard}, S. and {Wilby}, M.~J. and {Snik}, F. and {Lozi}, J. and {Guyon}, O. and {Norris}, B.~R.~M. and {Jovanovic}, N. and {Martinache}, F. and {Sauvage}, J. -F. and {Keller}, C.~U.},
        title = "{On-sky verification of Fast and Furious focal-plane wavefront sensing: Moving forward toward controlling the island effect at Subaru/SCExAO}",
      journal = {\aap},
         year = 2020,
        month = jul,
       volume = {639},
          eid = {A52},
        pages = {A52},
          doi = {10.1051/0004-6361/202037910},
archivePrefix = {arXiv},
       eprint = {2005.12097},
 primaryClass = {astro-ph.IM},
       adsurl = {https://ui.adsabs.harvard.edu/abs/2020A&A...639A..52B}
}

@ARTICLE{2023Bottom,
       author = {{Bottom}, Michael and {Walker}, Samuel A.~U. and {Cunnyngham}, Ian and {Guthery}, Charlotte and {Delorme}, Jacques-Robert},
        title = "{Sequential coronagraphic low-order wavefront control}",
      journal = {arXiv e-prints},
         year = 2023,
        month = dec,
          eid = {arXiv:2312.06806},
        pages = {arXiv:2312.06806},
          doi = {10.48550/arXiv.2312.06806},
archivePrefix = {arXiv},
       eprint = {2312.06806},
 primaryClass = {astro-ph.IM},
       adsurl = {https://ui.adsabs.harvard.edu/abs/2023arXiv231206806B}
}

@INPROCEEDINGS{2002Gonsalves,
       author = {{Gonsalves}, Robert A.},
        title = "{Adaptive Optics by Sequential Diversity Imaging}",
    booktitle = {ESO Conf. and Works. Proc.},
         year = 2002,
       editor = {{Vernet}, E. and {Ragazzoni}, R. and {Esposito}, S. and {Hubin}, N.},
       series = {ESO Conf. and Works. Proc.},
       volume = {58},
        month = jan,
        pages = {121},
       adsurl = {https://ui.adsabs.harvard.edu/abs/2002ESOC...58..121G}
}

@ARTICLE{2024Guiterrez,
       author = {{Gutierrez}, Yann and {Mazoyer}, Johan and {Mugnier}, Laurent M. and {Herscovici-Schiller}, Olivier and {Abeloos}, Baptiste},
        title = "{Image-based wavefront correction using model-free reinforcement learning}",
      journal = {Opt. Express},
         year = 2024,
        month = aug,
       volume = {32},
       number = {18},
        pages = {31247},
          doi = {10.1364/OE.529415},
archivePrefix = {arXiv},
       eprint = {2406.18143},
 primaryClass = {physics.optics},
       adsurl = {https://ui.adsabs.harvard.edu/abs/2024OExpr..3231247G}
}

@article{Korkiakoski:14,
author = {Visa Korkiakoski and Christoph U. Keller and Niek Doelman and Matthew Kenworthy and Gilles Otten and Michel Verhaegen},
journal = {Appl. Opt.},
number = {20},
pages = {4565--4579},
publisher = {Optica Publishing Group},
title = {Fast \& Furious focal-plane wavefront sensing},
volume = {53},
month = {Jul},
year = {2014},
url = {https://opg.optica.org/ao/abstract.cfm?URI=ao-53-20-4565},
doi = {10.1364/AO.53.004565},
}

@Article{Parvizi2023,
AUTHOR = {Parvizi, Payam and Zou, Runnan and Bellinger, Colin and Cheriton, Ross and Spinello, Davide},
TITLE = {Reinforcement Learning Environment for Wavefront Sensorless Adaptive Optics in Single-Mode Fiber Coupled Optical Satellite Communications Downlinks},
JOURNAL = {Photonics},
VOLUME = {10},
YEAR = {2023},
NUMBER = {12},
ARTICLE-NUMBER = {1371},
URL = {https://www.mdpi.com/2304-6732/10/12/1371},
ISSN = {2304-6732},
DOI = {10.3390/photonics10121371}
}

@ARTICLE{Huby+2015,
       author = {{Huby}, E. and {Baudoz}, P. and {Mawet}, D. and {Absil}, O.},
        title = "{Post-coronagraphic tip-tilt sensing for vortex phase masks: The QACITS technique}",
      journal = {\aap},
         year = 2015,
        month = dec,
       volume = {584},
          eid = {A74},
        pages = {A74},
          doi = {10.1051/0004-6361/201527102},
archivePrefix = {arXiv},
       eprint = {1509.06158},
 primaryClass = {astro-ph.IM},
       adsurl = {https://ui.adsabs.harvard.edu/abs/2015A&A...584A..74H}
}

\end{document}